\documentstyle[12pt]{article}

\def\reff#1{(\ref{#1})}
\newcounter{masection} 
\newcommand{\masection}[1]{\setcounter{equation}{0} 
\refstepcounter{masection} \vspace{10pt plus 10pt minus 3pt} \noindent 
{\large\bf \arabic{masection} #1}\par\vspace{5pt}} 
  
\renewcommand{\theequation}{\mbox{\arabic{masection}.\arabic{equation}}} 
 
\newcounter{masubsection}[masection] 
\newcommand{\masubsection}[1]{ \refstepcounter{masubsection} \vspace{5pt 
plus 5pt minus 2pt} \renewcommand{\themasubsection}{% 
\arabic{masection}.\arabic{masubsection}} \noindent {\bf 
\arabic{masection}.\arabic{masubsection} #1} \par\vspace{5pt}} 
 
\newcounter{saveeqn}

\newcommand{\ba}[1]{\begin{array}{#1}} 
\newcommand{\ea}{\end{array}} 
\newcommand{\be}{\begin{equation}} 
\newcommand{\ee}{\end{equation}} 
\newcommand{\bea}{\begin{eqnarray}} 
\newcommand{\eea}{\end{eqnarray}} 
\newcommand{\beann}{\begin{eqnarray*}} 
\newcommand{\eeann}{\end{eqnarray*}}

\newcommand{\D}{{\cal D}} 
\newcommand{\G}{{\cal G}} 
 
\newcommand{\R}{{\cal R}}

\newcommand{\pp}{{\prime\prime}}

 \newcommand{\tr}{{\rm tr}\,} 
\newcommand{\Alpha}{{p}} 

\newcommand{\Imm}{\mbox{\rm Im }} 
\newcommand{\Ree}{\mbox{\rm Re }}

\def\idty{{\leavevmode{\rm 1\ifmmode\mkern -5.4mu\else 
\kern -.3em\fi I}}} \def\Ibb #1{ {\rm I\ifmmode\mkern -3.6mu\else\kern 
-.2em\fi#1}} \def\Ird{{\hbox{\kern2pt\vbox{\hrule height0pt depth.4pt 
width5.7pt \hbox{\kern-1pt\sevensy\char"36\kern2pt\char"36} \vskip-.2pt 
\hrule height.4pt depth0pt width6pt}}}} 
\def\Irs{{\hbox{\kern2pt\vbox{\hrule height0pt depth.34pt width5pt 
\hbox{\kern-1pt\fivesy\char"36\kern1.6pt\char"36} \vskip -.1pt \hrule 
height .34 pt depth 0pt width 5.1 pt}}}} 
 \def\A1n{A_1\otimes\cdots\otimes A_n}  

  \def\ketbra 
\def\tr{\mathop{\rm tr}\nolimits} \def\Tr{\mathop{\rm Tr}\nolimits} 
 
\def\phi{\varphi}            % redefinition! 
\def\epsilon{\varepsilon}    % redefinition! 

\def\A{{\cal A}}   \def\D{{\cal D}} 
\def\G{{\cal G}}    
 \def\R{{\cal R}}   
  \def\I{{\cal I}}

\def\bbbc{{\mathchoice {\setbox0=\hbox{$\displaystyle    C$}\hbox{\hbox 
to0pt{\kern0.4\wd0\vrule height0.9\ht0\hss}\box0}} 
{\setbox0=\hbox{$\textstyle    C$}\hbox{\hbox 
to0pt{\kern0.4\wd0\vrule height0.9\ht0\hss}\box0}} 
{\setbox0=\hbox{$\scriptstyle    C$}\hbox{\hbox 
to0pt{\kern0.4\wd0\vrule height0.9\ht0\hss}\box0}} 
{\setbox0=\hbox{$\scriptscriptstyle    C$}\hbox{\hbox 
to0pt{\kern0.4\wd0\vrule height0.9\ht0\hss}\box0}}}} 
\def\bbbz{{\mathchoice 
{\hbox{$\sf\textstyle Z\kern-0.4em Z$}} {\hbox{$\sf\textstyle 
Z\kern-0.4em Z$}} {\hbox{$\sf\scriptstyle Z\kern-0.3em Z$}} 
{\hbox{$\sf\scriptscriptstyle Z\kern-0.2em Z$}}}} 
\def\ibb 
#1{\leavevmode\hbox{\kern.3em\vrule height 1.5ex depth -.1ex width 
.2pt\kern-.3em\rm#1}}  
  
\def\pp{{\Ibb P}}  
  
\def\Rl{{\Ibb R}} 
\def\qq{{{\bf{Q}}}} 
\def\rr{{{\bf{p}}}} 
\def\I{{\Ibb I}} 
\def\det{{\hbox{det}}} 
\def\str{{\hbox{Str}}}
\def\sdet{{\hbox{Sdet}}}
 
\def\eps{{\varepsilon}} 

\def\intsum{\mathop{\hbox to 0pt{$\sum$}\int}} 
 
\def\bS{{\bf{S}}} 
\def\cS{{\cal S}} 
\def\opsi{{\overline{\Psi}}} 
\def\ophi{{\overline{\phi}}} 
 
\def\ochi{{\overline{\chi}}} 
\def\olambda{{\overline{\Lambda}}}

\def\qh{\widehat{q}}
\def\wtq{\widetilde{q}}

\begin{document} 
%\renewcommand{\baselineskip}{1.5}

%  
%{\baselineskip=12pt \thispagestyle{empty} {Archived as
%{\tt cond-mat/9804008} and {\tt mp\_arc 98-267} \hspace{\fill}
%Preprint CPT-XXX-98}

\vspace{30pt}

\begin{center}
{\LARGE\bf Random Matrix approach to the crossover from Wigner to
Poisson statistics of energy levels}\\
%[30pt] 
\end{center}
\vspace{1truecm}
\begin{center}
\large{Nilanjana Datta and Herv\'e Kunz}
\end{center}
\vspace{.6truecm}
\begin{center}
Institut de Physique Th\'eorique
\end{center}
%\vspace{.2truecm}
\begin{center}
EPFL, CH-1015 Lausanne
\end{center}
%\vspace{.2truecm}
\begin{center}
Switzerland
\end{center}
\noindent
\begin{center}
{\bf Abstract}
\end{center}
\vspace{.2truecm}
We analyze a class of parametrized Random Matrix models, introduced by Rosenzweig and 
Porter, which is expected to describe the energy level statistics of quantum systems
whose classical dynamics varies from regular to chaotic as a function of a
parameter. We compute the
generating function for the correlations of energy levels, in the limit of infinite
matrix size. Our computations show that for a certain range of values of the parameter,
the energy-level statistics is given by that of the  
Wigner-Dyson ensemble. For another range of parameter values, one
obtains the Poisson statistics of uncorrelated energy levels. However, 
between these two ranges, new statistics emerge, which is neither 
Poissonnian nor Wigner. The crossover is
measured by a renormalized coupling constant. 
The model is exactly solved
in the sense that, in the limit of infinite matrix size, the energy-level
correlation functions and their
generating function are given in terms of a finite set of integrals.
\newpage
\masection{Introduction}
Random Matrix Theory (RMT) \cite{mehta}, originally introduced by Wigner, to characterize the 
statistical behaviour of the energy levels of nuclei, has found many successful
applications in various fields of physics in recent years.
Originally, it was thought that RMT was applicable
only to complex systems with many degrees of freedom. 
Hence, it came as a surprise 
when it was found that it could equally well describe simple quantum systems,
with very few degrees of freedom, as long as their classical dynamics
were chaotic. The first evidence of this fact was provided in the seminal
paper by Bohigas et al \cite{bohigas} in which the energy level fluctuations
of the quantum Sinai billiard were analysed and shown to be consistent 
with the predictions of the Gaussian Orthogonal ensemble of RMT.
Since this pioneering work, it has been checked numerically, on a wide
variety of systems, that the local statistical properties of a quantum 
system, whose classical counterpart is chaotic, are well-described by RMT.
In particular, the nearest neighbour spacing distribution was found to
be in excellent agreement with the spacing distribution beteen adjacent
eigenvalues of random matrices.

In contrast, Berry and Tabor \cite{berry} had 
given strong arguments to justify that, for integrable systems with more than
one degree of freedom, the  nearest neighbour spacing distribution of
the quantum energy levels should have a Poisson distribution, characteristic
of uncorrelated levels. This has been confirmed by many
numerical studies. There now exist excellent reviews on this topic [see 
\cite{bohigas2} and \cite{guhr1}].

However, it is well-known in classical mechanics that purely integrable
or purely chaotic systems are rare (at least for systems with a few degrees
of freedom). For most systems, the phase space is partitioned into
regular and chaotic regions and hence these systems can be referred to as 
{\em{mixed systems}}.

An important physical system illustrating these different behaviours,
is the hydrogen atom in a magnetic field. 
The classical system is essentially integrable
(chaotic) at weak (strong) fields but appears to be mixed at intermediate 
values of the field. This classical behaviour has its counterpart in the
energy level statistics of the corresponding quantum system, which exhibits 
a crossover from Poisson to Wigner type, when the magnetic field is 
increased \cite{delande}. It is, therefore, important to find models of 
random matrices which could 
describe the statistics of the energy levels of such mixed systems. 
Qualitatively, such a 
model should be governed by a Hamiltonian matrix which is essentially a sum 
of two parts, one describing the chaotic part of
phase space and hence belonging to the Wigner-Dyson ensemble of the 
relevant symmetry, and the other corresponding to
the regular part of phase space. A number of authors have studied models in 
which block-diagonal GOE matrices are weakly coupled by matrix elements
also belonging to a GOE ensemble \cite{pandey, leitner}. 
In this paper, we consider another
class of models, first introduced by Rosenzweig and Porter \cite{rose}
to describe the observed deviations from the Wigner- and Poisson statistics
in the spectra of some transition metal atoms.

This class of models is governed by an
ensemble of $N \times N$ matrices of the form
\be
H = A + \frac{\lambda}{N^\alpha} \, G
\label{haml}
\ee
where $A$ and $G$ are either real symmetric matrices (in the orthogonal
case) or complex self-adjoint ones (in the unitary case). The matrix elements
of $G$ are chosen to be independent random variables with a Gaussian 
distribution of unit variance. In the context of quantum chaos, $G$ is supposed
to correspond to the chaotic region of the classical phase space and
hence its eigenvalue distribution should obey the Wigner-Dyson statistics. 
In contrast, $A$  should correspond to the classically regular region 
and hence its eigenvalues should exhibit Poisson statistics. It is easily
seen that the statistics of the energy levels of $H$ depend only on 
the eigenvalues of $A$. Hence, without loss of generality, the matrix $A$ 
can be chosen to be diagonal. The simplest type of model which can be considered 
is, therefore, the following one: $A$ is a diagonal matrix, whose elements are 
independent random variables with a probability distribution $\nu (\cdot)$. 
Different behaviours can be expected by varying the exponent $\alpha$
in \reff{haml}.
%If $\alpha = 1/2$, one expects GUE (or GOE) statistics, whereas if $\alpha >1$, the
%statistics should be Poissonian. 
The case $\alpha = 1/2$, corresponds to a perturbed Wigner-Dyson ensemble.
It was recently analyzed by Brezin and Hikami \cite{brezin}.
They considered the case in which $G$ belongs to the Gaussian Unitary ensemble 
(GUE) and $A$ is a {\em{fixed}} diagonal matrix. They showed that the energy
level statistics for such a matrix ensemble 
was the same as that of $G$, i.e., the statistics relevant to the GUE. 
If $\alpha >1$, the energy level statistics is expected 
to be Poissonian.  
The value $\alpha=1$ corresponds to the 
{\em{crossover regime}} and for it one expects new statistics. In fact, by
making a numerical study of this model, Rosenzweig and Porter showed
that if one chooses the exponent $\alpha$ to be unity, then one obtains
energy level statistics which is intermediate between Wigner- and Poisson
statistics.  

Analytical studies of the model for $\alpha=1$, 
has been done only in the unitary case. 
These studies made use of certain special features of unitary
matrices. However, the case of the GOE, which one encounters more often, and
is technically more challenging, had remained
virtually unsolved thus far. The only results for this case were perturbative
ones in the small $\lambda$ limit \cite{leyv}.

In this article, we develop a technique which can be used to study the spectral
correlations for the case in which $G$ belongs to the GOE as well as to the 
one in which it belongs to the GUE. We compute
the generating function for the average value of the product of traces of 
advanced Green's functions, and the mixed product of traces of 
advanced and retarded Green's functions. All the correlation functions of
energy levels can be obtained from it, in the limit where $N$ goes to infinity.

In the case $\alpha=1/2$ we show that the statistics for correlations between energy 
levels, on the scale
of the mean level spacing, is the same as that of
the Wigner-Dyson Ensemble. In the unitary case, we hence 
recover the result of Brezin and Hikami \cite{brezin}. This matching of 
our result with theirs is not apriori obvious, because the mean level spacing 
depends on $A$ and we average over the distribution
of $A$, whereas they take it to be fixed.

In the case $\alpha=1$, our result for the generating function (in the infinite
$N$ limit) is in the form of a finite set of ordinary integrals. Quite 
generally, we show that the density of states at an energy $e$ 
is given by $\nu (e)$, and that 
all the correlation functions are universal functions depending only on
the ``renormalized'' coupling constant $\Lambda = \lambda \nu (e)$. This
suggests that in order to make a comparison of the results of
the model with a given quantum system, it might be plausible 
to take $\Lambda$ to be 
the ratio of $\rho (e)$ and $\rho_{\rm{reg}}(e)$, where $\rho (e)$ is the 
classsical Louiville measure of the energy surface and $\rho_{\rm{reg}}(e)$
is the measure corresponding to the regular part of the phase space.

In order to obtain more concrete results, one needs to evaluate the 
integrals appearing in the expression for the generating functions. 
This explicit computation, 
which turns out to be a rather lengthy one, has been done in this paper 
for the generating function for the two-point correlation function, in the unitary case. 
From it we can recover the two-point correlation function itself, in the
form of certain integrals over modified Bessel functions. In the orthogonal
case, the generating function, even for the two-point correlation function, 
appears in the form of integrals over elliptic functions and we postpone
the study of it to a future paper. 

It may be worth giving some hint about the technique used in this paper.
We basically use integrals over auxiliary Grassmannian variables to compute 
the average over the distribution of the Hamiltonian. However, finally, we
evaluate these Grassmannian integrals so as to arrive at a representation
in terms of ordinary integrals, in the large $N$ limit. Although similar in spirit 
to the familiar supersymmetric approach, introduced by Efetov \cite{efe} in this
kind of problems, our technique is different in that we never compute supersymmetric 
integrals (the only case in which we could find supersymmetry useful is when $\alpha = 1/2$).

\masection{Generating Function} 
We want to calculate the correlation functions $\rho^{(n)}(e_1, \ldots, e_n)$  
of the eigenvalues 
$\lambda_j$ of an $N\times N$ self--adjoint matrix $H$. They are defined as 
\be 
\rho^{(n)}(e_1, \ldots, e_n) = \langle \prod_{\alpha = 1}^n  
				\widehat{\rho} (e_\alpha)\rangle,  
\label{corr} 
\ee 
where  
\be 
\widehat{\rho} (e) = \frac{1}{N}\sum_j \delta(e - \lambda_j), 
\ee 
is the local density of eigenvalues at the energy $e$.  
The angular brackets will henceforth indicate an average over the probability 
distribution of $H$. 
 
If $G^{s}(e)$ denotes the  
advanced $(s = + 1)$ and retarded $(s=-1)$  
Green's function 
\be 
G^{s}(e) = \frac{1}{e - H - is\eps}, 
\label{green} 
\ee 
then  
\be 
\widehat{\rho} (e) = {\hbox{lim}\atop{\eps \uparrow 0^+}}  
	\left\{\frac{1}{2\pi i N} \tr \left[G^+(e) - G^-(e))\right]\right\}. 
\ee 
We will use the following identity to compute $[\tr\, G^s(e)]/{N}$. 
\be 
\frac{\partial}{\partial \eps_\pm} \, \frac{{\hbox{det}} \, 
[(\epsilon_-/N)\idty_N + 
i s \, (e\idty_N  - H)]}  
{{\hbox{det}} \,[(\epsilon_+/N)\idty_N  + 
i s \, (e\idty_N  - H)]} \biggr|_{\eps_- = \eps_+}  
= \mp \frac{1}{isN}\tr G^s(e). 
\ee
The symbol $\idty_N$ is used to denote 
the $N \times N$ identity matrix. 
Hence, it is evident that the correlation functions  
of energy levels can be obtained from the {\em{generating function}} 
\be 
J_n^S = \langle \prod_{\alpha=1}^n \frac{{\hbox{det}} \, 
[\epsilon_- (\alpha)/N + i s_\alpha \, (e_\alpha - H)]}  
{{\hbox{det}} \,[\epsilon_+ (\alpha)/N + 
i s_\alpha \, (e_\alpha - H)]} \rangle, 
\label{js1} 
\ee 
where $S= \{s_\alpha\}^n_{\alpha=1}$, $s_\alpha \in \{1, -1\}$ and  
 $\eps_\pm (\alpha) > 0 $, by taking suitable derivatives of it 
with respect to the variables 
$\eps_- (\alpha)$ or $\eps_+ (\alpha)$. 
 
In particular, the density of states is given by 
\be 
\rho(e) = \rho^{(1)}(e) = 
{\hbox{lim}\atop{\eps_- \uparrow 0^+}} 
{\hbox{Re}} \left\{\frac{1}{\pi}   
\frac{\partial}{\partial \eps_-}{ J^{+}_1}  
\biggr|_{\eps_- = \eps_+} \right\} 
\label{dos} 
\ee 
and the two--point correlation function by 
\be 
\rho^{(2)}(e_1, e_2) =  {\hbox{lim}\atop{\eps_- \uparrow  
0^+}} {\hbox{Re}} \left\{ 
\frac{1}{2 \pi^2} \frac{\partial^2 [J_2^{+-} +J_2^{++}] } 
{\partial \eps_-(1) \,\partial \eps_-(2)}  
\biggr|_{\eps_- = \eps_+}\right\}. 
\label{twopoint} 
\ee 
In this paper, we will consider Hamiltonians $H$ of the form 
\be 
H = A + \frac{\lambda}{N} G, 
\label{hgoe} 
\ee 
where $G$ is an $N \times N$ matrix whose elements are independent
random variables 
with a Gaussian distribution of unit variance and zero mean. When the matrix  
elements of $G$ are real (complex), the matrix $G$ belongs to the 
Gaussian orthogonal 
(unitary) ensemble of standard Random Matrix Theory. Since the probability  
distribution of $G$ is independent of the basis, the correlations of the 
energy  
levels of $H$ depend only on the eigenvalues $\{a_j\}$ of $A$. Hence, without
loss of generality, we can choose $A$ to be a diagonal matrix whose 
elements are independent random variables with a probability distribution 
$\nu (a_j)$.
%In the 
%sequel, we will therefore take for $A$ a diagonal matrix, whose elements 
%will be considered as independant random variables with a  
%common distribution $\nu (a)$. 
% 

\masubsection{The orthogonal case for finite matrix size} 
Consider a mixed system governed by a Hamiltonian matrix of  
the form \reff{hgoe}, with the matrix $G$ belonging  
to the Gaussian Orthogonal Ensemble ({\em{GOE}}). 
In this case, it is convenient to express the  
generating function $J^S$ as 
the ratio of the {\em{square roots}} of the determinants of an antisymmetric 
matrix and a symmetric one. This is because one can cast such a ratio as a  
product of integrals over real and Grassmannian variables, by making use of 
the integral identities \reff{id1} and \reff{id2}. 
[Note that here and henceforth, we suppress the subscript $n$ of the 
generating function $J^S_n$.] 
We can achieve this as follows: We start from the original expression  
\reff{js1} of the generating function, which can be rewritten in the form 
\be 
J^S = (\det \, S)^N \langle \frac{\det \, C_-}{\det \, C_+} \rangle, 
\label{js2} 
\ee 
with $C_+$ and $C_-$ being $nN \times nN$ matrices given by 
\be 
C_- = z_- \otimes \idty_N - i\,\idty_n \otimes H, 
\label{c-} 
\ee 
\be 
C_+ = z_+ \otimes \idty_N -  i\, S \otimes H. 
\ee 
Here $z_+, z_-$ 
and $S$ are $n \times n$ matrices with elements 
\bea 
z_- (\alpha \alpha') &=& \delta_{\alpha \alpha'}\, [ \frac{\eps_-(\alpha)}{N} s_\alpha 
+ i e_\alpha] \label{z-} \\ 
z_+ (\alpha \alpha') &=& \delta_{\alpha \alpha'}\, [ \frac{\eps_+(\alpha)}{N}  
+ i e_\alpha s_\alpha] \label{z+}\\ 
S(\alpha \alpha') &=& \delta_{\alpha \alpha'}\, s_\alpha. 
\label{defs} 
\eea 
The RHS of \reff{js2} can be cast in the form 
\be 
	(\det \,\cS)^{N/2} \, \langle  
\frac{\det \, M_-^{1/2}}{\det \, M_+^{1/2}} \rangle, 
\ee 
where $\cS = \idty_2 \otimes S$, and  
$M_+$ and $M_-$ are, respectively, the $2nN \times 2nN$ 
symmetric and antisymmetric matrices given by 
\be  
M_+ = \idty_2 \otimes C_+ \quad ; \quad M_- = \Upsilon \, [\idty_2 \otimes C_-], 
\ee  
with $\Upsilon$ being the $2n \times 2n$ antisymmetric matrix: 
$$ 
\Upsilon =  
\left( \begin{array}{lr} 0 & \idty_n \\
		  - \idty_n & 0 
\end{array} \right).
$$ 
This leads to the expression  
\be 
J^S = \bigl( \det \, \cS)^{N/2} \, \langle \left[\frac{\det \, (\Upsilon Z_- \otimes \idty_N 
 - i \Upsilon \otimes H)}{\det \, ( Z_+ \otimes \idty_N 
 - i \cS \otimes H)}\right]^{1/2}\rangle, 
\label{js3} 
\ee 
where the matrices, $Z_\pm = \idty_2 \otimes z_\pm$, have the matrix elements  
\be 
Z_\pm(p\alpha | p' \alpha') = \delta_{\alpha \alpha'} \, \delta_{p p'} z_\pm  
(\alpha \alpha') , 
\label{zpm} 
\ee 
with $p = 1,\, 2$. Moreover, 
\bea 
\cS(p\alpha | p' \alpha') &=& \delta_{\alpha \alpha'} \, \delta_{p p'}  
s_\alpha,\\ 
\Upsilon(p\alpha | p' \alpha') &=&  \delta_{\alpha \alpha'} \, \gamma{(p p')}, 
\eea 
where  
\be 
\gamma(pp) = 0 \quad; \quad \gamma(12) = 1; \quad; \quad  
\gamma(21) = -1. 
\label{gamma} 
\ee 
We now use the standard trick of expressing the square root of 
the determinant appearing in the numerator of the expression 
\reff{js3} as a Gaussian integral over Grassmannian variables, 
and for the one appearing in the denominator, a usual Gaussian 
integral over real variables. [See Appendix A for a summary of  
some useful identities]. In this way the generating function  
can be written as a superintegral.  
\be 
J^S = (\det \, \cS)^{N/2} \int D\, \Phi \, \langle e^{-\Bigl(\Phi,  
\bigl(\Gamma Z  
\otimes \idty_N - \Gamma \bS \otimes H\bigr) \Phi\Bigr)} \rangle, 
\label{two2}
\ee 
where $\Gamma, Z$ and $\bS$ are $4n \times 4n$ matrices defined as follows:\\ 
For $\sigma = + , -\,$;  
$$ Z = Z_\sigma \, \delta_{\sigma \sigma'} \, \, ;\, \, 
\bS = \bS_\sigma \, \delta_{\sigma \sigma'} \, \, ;\, \, 
\Gamma = \Upsilon_\sigma \, \delta_{\sigma \sigma'}, 
$$  
with  
\be 
\Upsilon_+ = \idty_{2n}, \, \Upsilon_- = \Upsilon, \, \bS_+ = \cS,  
\bS_- = \idty_{2n}; 
\ee 
and $\Phi$ denotes a supervector  
$$ 
\Phi =  
\left( \begin{array}{c} \phi_+ \\ 
		   \phi_-  
\end{array} \right), 
$$ 
i.e., $\phi_+$ and $\phi_-$ are, respectively, vectors with real 
and Grassmannian elements, labeled by  
$$ 
\phi_{\pm i} (p \alpha) \quad {\hbox{with}} \quad i \in \{1,\ldots, N\}, 
\, p\in \{1,2\}, \, \alpha \in \{1, \ldots n\}. 
$$ 
At this point, it is important to note a basic difference between the
sector of real variables corresponding to $\phi_+$, and the Grassmannian 
sector corresponding to 
$\phi_-$. If in \reff{z-} and \reff{z+}, all the energies $e_\alpha$ 
are the same and $\eps_\alpha=0$ for all $\alpha \in \{1 \ldots n\}$, 
then the integrand in
\reff{two2} as well as the measure $D \Phi$ remain invariant 
under the following change of variables:
\be
\phi_{+i}(p \alpha) = \sum_{p',\alpha'} \tau(p\alpha|p'\alpha') \, 
\phi_{+i}^\prime(p'\alpha'),
\ee 
where $\tau$ is a $2n \times 2n$ matrix satisfying the relation
\be
\tau^t \cS \tau = \cS;
\label{hyp}
\ee
and
\be
\phi_{-i}(p \alpha) = \sum_{p',\alpha'} U(p\alpha|p'\alpha') \, 
\phi_{-i}^\prime(p'\alpha'),
\ee 
where $U$ is a $2n \times 2n$ matrix such that
\be
U^t \Upsilon U = \Upsilon.
\label{ortho}
\ee
The matrices $\tau$ form a group. If there are
$p$ variables $s_\alpha$ taking the value $+1$ and $q$ variables taking
the value $-1$, then this group is $O(p,q)$. Unless $q=0$ or $p=0$, 
this is a non-compact
hyperbolic group. Since the sign of $s_\alpha$ corresponds to the
type of the Green's function (i.e., {\em{advanced}} or {\em{retarded}}),
it is evident  that this hyperbolic symmetry comes into play when we compute
the average of a product of $p$ advanced and $q$ retarded Green's functions.
In contrast, the matrices $U$ belong to the $2n$ symplectic group. 
We have chosen to extract the factors of $s_\alpha$ in the 
numerator of the expression \reff{js3}, of the generating function, in 
order to emphasize the difference between these two sectors.

Using the expression, \reff{hgoe}, of the mixed Hamiltonian $H$, 
we obtain 
\bea 
J^S &=& (\det \, \cS)^{N/2} \int D\, \Phi \, \int D\nu(A) \, 
\exp\left[ - \Bigl(\Phi, \bigl(\Gamma Z \otimes \idty_N - 
\Gamma \bS \otimes A \bigr)  
\Phi\Bigr)\right] \nonumber\\ 
& & \quad \times \quad \langle \exp (\Phi, \bigl(i \frac{\lambda}{N} \Gamma \bS   
\otimes G \bigr) \Phi\Bigr) \rangle, 
\label{jay1} 
\eea 
where  
\be 
D\nu(A) := \prod_{j=1}^N da_j \, \nu(a_j) 
\label{measure} 
\ee 
In \reff{jay1} the angular brackets denote an averaging over the Gaussian  
Orthogonal ensemble.\\ 
By integrating over the Gaussian probability distribution  
of the matrix elements of $G$, we see that 
\be 
\langle \exp (\Phi, \bigl(i \frac{\lambda}{N} \Gamma \bS   
\otimes G \bigr) \Phi\Bigr) \rangle =  
\exp \left( - \frac{\lambda^2}{2N^2} \, \tr Y^2\right), 
\label{tra2} 
\ee 
where $Y$ is the $2n \times 2n$ matrix defined as follows: 
\be 
Y = \sum_{\sigma = +, -} Y_\sigma, 
\label{Y} 
\ee 
with 
\be 
Y_\sigma(ij) = Y_\sigma(ji) = \sum_{p, p' = 1} \sum_{\alpha=1}^{n} 
\phi_{\sigma i}(p \alpha) \, \phi_{\sigma j}(p' \alpha)  \Upsilon_\sigma(p p') 
\bS_\sigma(\alpha). 
\ee 
Hence, 
\bea 
J^S &=& (\det \, \cS)^{N/2} \int D\, \phi_+ \,\int D\, \phi_-  \, \int D\nu(A) \, 
\exp \left[- \Bigl(\phi_+, \bigl(Z_+\otimes \idty_N - i \cS \otimes A \bigr)  
\phi_+\Bigr)\right]\nonumber\\ 
& & \exp \left[- \Bigl(\phi_-, \bigl(\Upsilon_- Z_-\otimes \idty_N - i  
\Upsilon_- 
\otimes A \bigr)  
\phi_-\Bigr)\right] \, e^{-\frac{\lambda^2}{2N^2} \tr Y^2} 
\label{jay11} 
\eea 
Since $\lambda$ appears on the RHS of \reff{tra2} only in the 
form of $\lambda^2$, it follows that we can always choose $\lambda$ 
to be positive. 
We shall make this choice for the rest of the paper. 
 
We can now decompose $\tr Y^2$ in the following way: 
\be 
\tr Y^2 = \tr (L_+ \cS)^2 - \tr (L_- \Upsilon_-)^2 +  
2 \left( \phi_-, \bigl( \Upsilon_- \otimes T\bigr) \phi_-\right), 
\ee 
where the $2n \times 2n$ matrices $L_\sigma$ are given by 
\be 
L_\sigma(p\alpha|p' \alpha') =  \sum_i \phi_{\sigma i}(p \alpha)\, 
\phi_{\sigma i}(p' \alpha') 
\ee 
and the $N \times N$ matrix $T$ is given by 
\be 
T_{ij} = \sum_{p \alpha} \phi_{+i}(p \alpha)\,\phi_{+j}(p \alpha) s_\alpha. 
\label{tee} 
\ee 
%%%%%%%% INSERTION %%%%%%%%%%%%%%%%%%%%%%%%%%%%%%%%%%%%%%%%%%%%%%%%%%%%%% 
%%%%%%%%%%%%%%% begin insertion %%%%%%%%%%%%%%%%%%%% 
We will consider correlations between energy levels around some 
energy $e$, on the scale of the mean level spacing, i.e., $1/(N\rho(e))$. 
For this purpose we decompose the energies $e_\alpha$ into 
\be 
e_\alpha = e + \frac{r_\alpha}{N} 
\ee 
so that if we define the matrices $\rho_+$ and $\rho_-$ of elements
\bea 
\rho_+(p\alpha|p'\alpha')&=& \delta_{\alpha \alpha'} \delta_{p p'}  
			\left[ \eps_+(\alpha) + i r_\alpha s_\alpha\right], 
\label{rho+}\\ 
\rho_-(p\alpha|p'\alpha')&=& \delta_{\alpha \alpha'} \delta_{p p'}  
			\left[ \eps_-(\alpha)s_\alpha + i r_\alpha \right], 
\label{rho-} 
\eea 
then the generating function can be written as 
\bea
J^S &=& (\det \, \cS)^{N/2}\,  e^{\frac{1}{2 \lambda^2}  
\left[\tr(\rho_+ \cS)^2 - \tr \rho_-^2\right]}
%\nonumber\\ & & \times 
\, \int D\nu(a)  
	\int D\phi_+ \,\int D\phi_- \,  
 e^{- \frac{\lambda^2}{2N^2} \tr [L_+^t \cS)^2 +  
\frac{\rho_+ \cS N}{\lambda^2}]^2} \nonumber\\ 
& &   \times \, 
e^{\frac{\lambda^2}{2N^2}  
\tr [L_-^t\Upsilon_- - \frac{\rho_-N}{\lambda^2}]^2}
\, e^ {- \Bigl(\phi_+, \bigl(i \cS \otimes (e\idty_N -  A \bigr)  
\phi_+\Bigr)}  
\,e^{-\Bigl(\phi_-, \bigl(i \Upsilon_- \otimes \frac{\lambda^2}{N^2} T\bigr) 
\phi_- \Bigr)} \nonumber\\ 
\label{jay2} 
\eea 
%%%%%%%% END INSERTION %%%%%%%%%%%%%%%%%%%%%%%%%%%%%%%%%%%%%%%%%%%%%%%%%% 
% 

%%%%%%%%%%%%%end insertion %%%%%%%%%%%%%%%%%%% 
\masubsection{Pseudo--Gaussian transformations}  
\label{transf} 
With the help of Gaussian integrations over auxiliary matrices $Q_+$ and 
$Q_-$, the exponents in the integrands [on the RHS of \reff{jay2}] can 
be reduced to quadratic forms in $\phi_+$ and $\phi_-$. This procedure, 
which is analogous to the Hubbard Stratonovich transformation, yields the  
following identities: 
\be 
e^{- \frac{\lambda^2}{2N^2}  \tr (L_+^t\cS + \frac{N}{\lambda^2}\cS \rho_+)^2} 
= \frac{1}{d_+} \, \int DQ_+\, e^{-\frac{1}{2} \tr Q_+^2}\, 
e^{- \frac{i \lambda}{N} \tr \left[Q_+ \left(L_+^t \cS + \frac{N}{\lambda^2} \cS  
\rho_+\right)\right]}, 
\ee 
and 
\be 
e^{ \frac{\lambda^2}{2N^2}  \tr (L_-^t\Upsilon -  
\frac{N}{\lambda^2}\rho_-)^2} 
= \frac{1}{d_-} \, \int DQ_-\, e^{-\frac{1}{2} \tr Q_-^2}\, 
e^{- \frac{\lambda}{N} \tr \left[Q_- \left(L_-^t \Upsilon_- -  
\frac{N}{\lambda^2} \rho_-\right)\right]}, 
\label{q-1} 
\ee 
with 
\be
d_\pm = \int DQ_\pm\, e^{-\frac{1}{2} \tr Q_\pm^2}.
\label{dpm}
\ee
\smallskip 
 
Our aim is to substitute the above identities on the RHS of \reff{jay2} 
and change the order of integration over $\phi_+, \phi_-$ and $Q_+, Q_-$. 
This is because the resulting integrals over $\phi_+$ and $\phi_-$ turn 
out to be Gaussian integrals, whose values are obtained by making 
use of the integral identities \reff{id1} and \reff{id2}. However, the 
change in the order of the integrations imposes a restriction on the 
form of the matrix $Q_+$. It is required to be of the following form 
[as introduced in \cite{schafer}]: 
\be 
Q_+ =  
\left( \begin{array}{lr} Q_{11} - i \sqrt{\delta^2 +  Q_{12}\, Q_{21}} & 
Q_{12} \\ 
		  Q_{21} &  Q_{22} + i \sqrt{\delta^2 +  Q_{21}\, Q_{12}} 
\end{array} \right). 
\label{q+} 
\ee 
%with $Q_{ij}$, $i,j \in {1,2}$ being real matrices, such that $Q_{11}$ 
%and $Q_{22}$ are symmetric, while $  Q_{21}= \left( Q_{12}\right)^t$; 
%$\delta$ is an arbitrary constant. For $n = 1$, the matrices $Q_+$ 
%reduce to complex numbers of the form 
%\be 
%q_+ = q'_+ - i \delta, \quad {\hbox{with}} \quad q'_+ \in \Rl \quad  
%{\hbox{and}} \delta > 0. 
%\label{delta} 
%\ee 
%The above form \reff{delta} is imposed by the requirement that the  
%integrals over $\phi_+$ and $q_+$ can be interchanged.  
%More generally, when $S$ is a multiple 
%of the identity matrix $\idty_n$ (i.e., $s_\alpha = s$ for 
%all $\alpha = 1 \ldots n$, with $s \in \{-1, 1\}$)  
%the matrices $Q_+$ are restricted to be of the form 
%\be 
%Q_+ = Q'_+ - i \delta \, \idty_{2n}, 
%\label{all} 
%\ee 
%with $\delta >0$ and $Q'_+$ being a real symmetric matrix.  
The block structure refers to the decomposition of the diagonal matrix $S$
[\reff{defs}]  
into $p$ elements equal to $+1$ and $q$ elements equal to $-1$, 
so that $Q_{11}$ is a  
$2np \times 2np$ real symmetric matrix, $Q_{22}$ a $2nq \times 2nq$ real 
symmetric  
matrix, $Q_{12}$ is a $2nq \times 2np$ real matrix and the $2np \times 
2nq$ matrix $Q_{21}$ satisfies 
\be 
Q_{21}^t = Q_{12}, 
\ee  
where the superscript $t$ denotes the transpose of the matrix. The variable 
$\delta$ appearing on the RHS of \reff{q+} denotes an arbitrary 
positive number. 
This structure of the matrix $Q_+$ is dictated by the invariance
of the integral over $\phi_+$ (on the RHS of \reff{jay2}) under
the pseudo-orthogonal group $O(p,q)$, where $p$ denotes the number
of advanced Green's functions and $q$ denotes the number of
retarded Green's functions.
\smallskip

We consider $Q_-$ to be Hermitian : $Q_-^\dagger = Q_-$, so for $n=1$ 
it reduces to a real number.  
In addition, we require $Q_-$ to satisfy the following relation: 
\be 
\left(Q_- \Upsilon_-\right)^t = - \left(Q_- \Upsilon_-\right) 
\label{q-} 
\ee 
This constraint is imposed so that the Gaussian  
integration over the Grassmannian variables 
$\{\phi_{-j}\}$ can be expressed in terms of the determinant of a matrix,  
as in \reff{id2} [see \reff{ei} -- \reff{rrr} below].  
 
The constraint \reff{q-} on the matrix $Q_-$ is imposed so as to ensure
the validity of the identity \reff{q-1}. Indeed the latter identity,
involving such a matrix $Q_-$, uses the fact that 
the linear term in $Q_-$, appearing in the exponential, is of the 
form $\tr(Q_- E \Upsilon_-)$, with $E$ being an antisymmetric matrix. 
We can therefore write the generating function in the form 
\bea 
J^S &=& \frac{(\det \, \cS)^{N/2}}{d_+\, d_-}\,  
 \int DQ_+ \int DQ_- \, \exp\left(-\frac{1}{2} \tr\Bigl[Q_+ + i 
\frac{\cS \rho_+}{\lambda}\Bigr]\right)^2\nonumber\\ 
& & \quad \times\, \exp\left(-\frac{1}{2} \tr\Bigl[Q_- -  
\frac{\rho_-}{\lambda}\Bigr]\right)^2 \, \widehat{K}(Q_+, Q_-)\nonumber\\ 
\label{jay4} 
\eea 
with 
\be 
\widehat{K}(Q_+, Q_-)  = \int D\nu(a) \, \int D\phi_+ \, 
e^ {- \Bigl(\phi_+, \bigl(i \cS \otimes (e\idty_N -  A \bigr)  
\phi_+\Bigr)} \, e^{-\left(\phi_+, (\frac{i\lambda}{N}\cS Q_+ \otimes  
\idty_N) \phi_+\right)} \times \I_-, 
\label{khat} 
\ee 
where $\I_-$ is the integral over the Grassmannian variables $\{\phi_{-j}\}$ 
and is given by 
\be 
\I_- = \int \D \phi_- \, \exp \left[ -(\phi_-, R\phi_-) \right] ,
\label{ei} 
\ee 
with 
\be 
R := \frac{\lambda}{N} (\Upsilon_- Q_-) \otimes \idty_N +  
\Upsilon_- \otimes \Bigl( i (e\idty_N -  A ) + \frac{\lambda^2}{N^2}  
T\Bigr). 
\label{ar} 
\ee 
The matrix $T$ is defined by \reff{tee}. 
	In the following section we evaluate the integral 
$\I_-$ and show that $\widehat{K}(Q_+, Q_-)$ depends only on the eigenvalues 
of the matrices $Q_+$ and $Q_-$.   
%%%%%%%%%%%%%%%%%%%%%%%%%%%%%%%%%%%%%%%%%%%%5 

\masubsection{Evaluation of the integral $\I_-$} 
 
Note that the matrix $R$, defined by \reff{ar}, is antisymmetric. This follows from our 
choice [\reff{q-}] of $(Q_- \Upsilon_- )$ to be an antisymmetric matrix.  
Hence, using the Gaussian identity, \reff{id2}, we obtain 
\be 
\I_- = \sqrt{\det R}. 
\label{rrr} 
\ee 
In fact, the constraint \reff{q-} on the matrix $Q_-$ was imposed 
so as to obtain the above result. 
\smallskip 
 
Moreover, we can show, as follows, that the matrix $R$ depends only on the  
{\em{eigenvalues}} of the matrix $Q_-$: Defining an $N\times N$ matrix $A_1$:  
\be A_1 := i (e\idty_N -  A ) + \frac{\lambda^2}{N^2}  
T, \label{a1} 
\ee 
we write 
\bea 
\det\, R &=& \det\left(\Upsilon_- \otimes \idty_N\right) \,  
\det \left[ \frac{\lambda}{N} Q_- \otimes \idty_N + \idty_{2n} 
\otimes A_1\right], \nonumber\\ 
&=& \det\left(\Upsilon_- \otimes \idty_N\right) \,  
\det \left[ \frac{\lambda}{N} q_- \otimes \idty_N + \idty_{2n} 
\otimes A_1\right] 
\nonumber\\ 
&=:& \det\, C, 
\label{see} 
\eea 
where $q_-$ is the diagonal matrix whose diagonal elements are  
the eigenvalues 
of $Q_-$. 
\smallskip 
 
The elements of the matrix $C$, \reff{see}, are given by 
\be 
C(p \alpha, j| p' \alpha', j') = \Upsilon_-(pp') \left[ 
\frac{\lambda}{N} q_-(\alpha) \delta_{\alpha \alpha'} \, \delta_{jj'}  
+ \delta_{\alpha \alpha'} A_1(jj')\right], 
\ee 
where $q_-(\alpha) \equiv q_-(p \alpha)$.  
This follows from the fact that the eigenvalues of $Q_-$ are doubly degenerate 
(as shown in Appendix B). 
Hence, the matrix $C$ is {\em{antisymmetric in the label}} $p$ and has the  
form 
$$ 
C =  
\left( \begin{array}{lr} 0 & D \\ 
		  - D & 0  
\end{array} \right) 
$$ 
where $D$ is an $nN \times nN$ matrix defined as follows:  
\be 
D:= \frac{\lambda}{N} \wtq_- \otimes \idty_N + \idty_{n} 
\otimes A_1. 
\label{dee} 
\ee 
In \reff{dee}, $\wtq_-$ denotes an $n \times n$ diagonal matrix with 
diagonal elements $\wtq_{-\alpha} = q_{-\alpha}$. 
This leads to the result 
\be 
\I_- = \det D 
\label{id} 
\ee 
From \reff{a1} and \reff{dee} it follows that 
\be  
\det D = \prod_{\alpha=1}^n \det\, \left( 
\wtq_{-\alpha} \frac{\lambda}{N}\, \idty_N + i(e\idty_N -A) + \frac{\lambda^2}{N^2}T \right). 
\label{dee2} 
\ee 
For each $\alpha\in \{1, \ldots n \}$, let us define an $N \times N$  
diagonal matrix  
\be 
b_\alpha := \wtq_{-\alpha} \frac{\lambda}{N}\, \idty_N + i(e\idty_N - A)  
\label{bee} 
\ee 
In terms of this matrix, we can write 
\be 
\I_- \equiv \det D = \prod_{\alpha} \det \,b_\alpha \, \det \left( \idty_N +  
\frac{\lambda^2}{N^2}b_\alpha^{-1}T\right) 
\label{dee3} 
\ee 
For each $\alpha$, let $F_\alpha$ denote a $2n \times 2n$ matrix whose  
elements are given by 
\be 
F_\alpha(p \gamma| p' \gamma') := \sum_{j=1}^N \left(b_\alpha^{-1}\right)_j\, 
\phi_{+j}(p \gamma)\,\phi_{+j}(p' \gamma') 
\label{tau} 
\ee 
It is easy to see that  
\be 
\tr\left(b_\alpha^{-1} T\right)^j = \tr\left(\cS F_\alpha\right)^j 
\label{trb} 
\ee 
for any arbitrary integer $j$. Hence by \reff{det3} we have the identity 
\be 
\det \left( \idty_N +  
\frac{\lambda^2}{N^2}b_\alpha^{-1}T\right) = 
\det \left( \idty_{2n} +  
\frac{\lambda^2}{N^2}\cS F_\alpha\right) 
\label{equiv} 
\ee 
Hence from \reff{dee3}, \reff{bee} and \reff{equiv} it follows that 
\be 
\I_- = \prod_{j=1}^N \det\left( \frac{\lambda}{N} \, \wtq_- + i(e - a_j) 
\idty_n\right) 
\, \prod_{\alpha=1}^n \det \left(\idty_{2n} +  
\frac{\lambda^2}{N^2}\cS F_\alpha\right) 
\label{ei2} 
\ee 
where $a_j$ denotes a diagonal element of the diagonal matrix $A$. 
Using the Grassmannian integral representation \reff{id3} for a  
determinant we can write 
\be 
\prod_{\alpha} \det\,\left(\idty_{2n} + \frac{\lambda^2}{N^2}\cS F_\alpha\right) 
=  \int D{\overline{\Psi}} \, D\Psi \, e^{-({\overline{\Psi}}, \Psi)} \, 
e^{-({\overline{\Psi}, B \Psi})}, 
\label{grass} 
\ee 
where $\Psi$(${\overline{\Psi}}$) is a column (row) vector of length $2n^2$, 
and  
\be 
({\overline{\Psi}, B \Psi}) = \sum_{\alpha=1}^{n} \sum_{\beta,\beta'=1}^{2n} 
\opsi_\alpha(\beta) B_\alpha(\beta\beta') \Psi_\alpha(\beta'), 
\ee 
where $\beta$ now refers to the double index $(p \gamma)$ and 
$B_\alpha$ denotes a $2n\times 2n$ matrix whose elements are given by 
\be 
B_\alpha(\beta \beta') = \frac{\lambda^2}{N^2} \cS (\beta) \,  
F_\alpha(\beta \beta'),
\ee 
with $\cS(\beta) = s_\gamma$.
The relations \reff{ei2} and \reff{grass} yield the following  
expression for the integral $\I_-$: 
\bea 
\I_- &=& \int D{\overline{\Psi}} \, D\Psi \, e^{-({\overline{\Psi}}, \Psi)} \, 
\prod_{j=1}^N \det\left( \frac{\lambda}{N} \, \wtq_- + i(e - a_j)\idty_n\right) 
\nonumber\\ 
& &\exp\left(-\Bigl[\frac{\lambda^2}{N^2} \sum_{j=1}^N \, \sum_{\beta, \beta'=1}^{2n} 
\phi_{+j}(\beta)\,\phi_{+j}(\beta')\,\cS_\beta\, \sum_{\alpha=1}^{n}  
%\left[i(e\idty_N - A) + \frac{\lambda}{N} q_{-a}\right]^{-1}_j  
\left[b_\alpha^{-1}\right]_j \,\opsi_a(\beta) \, \Psi_a(\beta')\Bigr]\right).
\nonumber\\ 
\label{ei3} 
\eea 
We can now insert the representation of $\I_-$, given by \reff{ei3},
in the expression \reff{khat} for $\widehat{K}$, and perform the integration
over $\phi_+$. This yields
\bea 
\widehat{K}(Q_+, Q_-)  &=& \int D{\overline{\Psi}} \, D\Psi \,  
e^{-({\overline{\Psi}}, \Psi)} 
 \,\int D\nu(a) \,\left[ \prod_{j'=1}^N \det\left( \frac{\lambda}{N} \, \wtq_- +  
i(e - a_{j'})\idty_{n}\right) \right]\nonumber\\ 
& & \prod_{j=1}^{N} \det \left[ \cS\left( i(e - a_{j})\idty_{2n} +  
i \frac{\lambda}{N} Q_+ 
+ \frac{\lambda^2}{N^2} \R_{j}\right) \right]^{-\frac{1}{2}}, 
\eea 
where each $R_j$ is a $2n \times 2n$ matrix with elements
\bea 
\R_j(\beta\beta') &=& \sum_{\alpha=1}^{n} \left( b^{-1}_{\alpha}\right)_j  
\opsi_{\alpha}(\beta)\,\Psi_{\alpha}(\beta')\nonumber\\ 
&=& \sum_{\alpha=1}^{n} \left( \frac{\lambda}{N} \wtq_{-\alpha}\idty_N + 
i (e\idty_N - A)\right)^{-1}_j 
\opsi_{\alpha}(\beta)\,\Psi_{\alpha}(\beta') 
\label{arj} 
\eea 
\smallskip 
From the definition \reff{measure} of the measure $D\nu(a)$, it follows that  
the expression for $\widehat{K}(Q_+, Q_-)$ involves a product of $N$ identical 
integrals and hence can be written in the form: 
\be 
\widehat{K}(Q_+, Q_-)  = (\det \cS)^{-N/2} \, 
\int D{\overline{\Psi}} \, D\Psi \,  
e^{-({\overline{\Psi}}, \Psi)} \, Z^N, 
\label{last} 
\ee 
where, writing $a$ for $a_j$ and $\R$ for $\R_j$, we define 
\be 
Z := \int da \, \nu(a)   
\det\left( \frac{\lambda}{N} \, \wtq_- +  
i(e - a)\idty_n \right) \, \det \left[ \cS\left( i(e - a)\idty_{2n} +  
i \frac{\lambda}{N} Q_+ 
+ \frac{\lambda^2}{N^2} \R \right) \right]^{-\frac{1}{2}}. 
\label{pee} 
\ee
%%%% 
That this expression depends only on the eigenvalues $q_+(p\alpha)$ of 
$Q_+$ can be easily seen by making the following change of 
variables in the Grassmannian variables 
\bea 
\ochi_\alpha(\beta) &=& \sum_{\beta^{''}} \G^{-1}(\beta \beta^{''})  
\opsi_{\alpha}(\beta^{''})\\ 
\chi_\alpha(\beta') &=& \sum_{\widetilde{\beta}} \G^{t}(\beta' 
\widetilde{\beta}) \Psi_{\alpha}(\widetilde{\beta}), 
\eea 
where $\G$ is the matrix which diagonalizes $Q_+$, i.e.,  
\be
Q_+ = \G q_+ \G^{-1},
\label{qg}
\ee 
and $q_+$ is a $2n \times 2n$ diagonal matrix with diagonal elements 
$q_+(p \alpha)$. 
Hence $\widehat{K}$ takes the form 
\be 
{\widehat{K}} = (\det \cS )^{-N/2} \int D \ochi D \chi e^{-(\ochi, \chi)} P^N 
\label{hk} 
\ee 
where 
\be 
P:= \int da \, \nu(a)\, \det\left( \frac{\lambda}{N} \, \wtq_{-} +  
i(e - a)\idty_n \right)  \, \det\left[ i(e - a)\idty_{2n} +  
i \frac{\lambda}{N} q_+ 
+ \frac{\lambda^2}{N^2} \R \right]^{-1/2}, 
\label{pee2} 
\ee 
with
\be 
\R(\beta \beta') = \sum_{\alpha = 1}^n \left( \frac{\lambda}{N} \wtq_{-\alpha}
 +  
i(e -a )\right)^{-1}\, \ochi_\alpha(\beta)\,\chi_\alpha(\beta').
\label{arr}
\ee 
%%%% 
\smallskip

We will cast $P$ in a slightly simpler form by making use of the degeneracy  
of the eigenvalues of $Q_-$. Let $X_\pm$ be the $2n\times 2n$  
diagonal matrices of diagonal elements 
\be 
x_+(p \gamma) := i(e - a) + i \frac{\lambda}{N} q_+(p\gamma) ; 
\label{x+} 
\ee 
and 
\be 
x_- (p \gamma)\equiv x_-(\gamma)  := i(e - a) +  \frac{\lambda}{N} q_-(\gamma). 
\label{x-} 
\ee 
We can rewrite $P$ [\reff{pee2}] as 
\be 
P= \int da \, \nu(a)\, \left[\frac{\det X_-}{\det X_+}\right]^{1/2} 
\, \det\left( \idty_{2n} + \frac{\lambda^2}{N^2}{X_+}^{-1} {\R} \right)^{-1/2}. 
\label{pee3} 
\ee 
The matrix elements of $(X_+^{-1} \R)$ are 
\be 
(X_+^{-1} \R) (p \gamma|p'\gamma') =\frac{1}{x_+(p\gamma)}  
\sum_{\alpha = 1}^{n} 
\frac{\ochi_\alpha(p \gamma) \,  
\chi_\alpha(p \gamma)}{x_-(\alpha)}. 
\ee 
It is easy to see that {{for any arbitrary integer $j$}} 
\be 
\tr  \left(X_+^{-1} \R\right)^j = - \tr \left(X_-^{-1} M \right)^j, 
\ee 
where $M$ is the $n \times n$ matrix of elements 
\be 
M_{\alpha_1 \alpha_2}(q_+) = \sum_{p=1}^2 \sum_{ \gamma=1}^n 
\frac{\chi_{\alpha_1}(p \gamma) \,\ochi_{\alpha_2}(p \gamma)}{X_+(p \gamma)}, 
\ee 
and $x_-$ is a diagonal $n \times n$ matrix of elements $x_-(\gamma)$. 
We can therefore write $P$ as 
\be 
P= \int da \, \nu(a)\, \left[\frac{\det X_-}{\det X_+}\right]^{1/2} 
\, \det\left( \idty_{n} +  \frac{\lambda^2}{N^2} \frac{M(X_+)}{x_-}  
\right)^{1/2}. 
\label{pee5} 
\ee 
Hence, the multiple integral $\widehat{K}(Q_+, Q_-)$, defined through  
\reff{khat}, is seen to depend only on the eigenvalues of the matrices 
$Q_+$ and $Q_-$. It is given by \reff{hk}, with $P$ being given  
by \reff{pee5}, above. 

Hence, the generating function is given by 
\bea 
J^S &=& \frac{(\det \, \cS)^{N/2}}{d_+\, d_-}\,  
 \int DQ_+ \int DQ_- \, \exp\left(-\frac{1}{2} \tr\Bigl[Q_+ + i 
\frac{\cS \rho_+}{\lambda}\Bigr]\right)^2\nonumber\\ 
& & \quad \times\, \exp\left(-\frac{1}{2} \tr\Bigl[Q_- -  
\frac{\rho_-}{\lambda}\Bigr]\right)^2 \, \widehat{K}(Q_+, Q_-),
\eea 
where
\be 
{\widehat{K}} = (\det \cS )^{-N/2} \int D \ochi D \chi e^{-(\ochi, \chi)} P^N,
\ee 
and $P$ is given by \reff{pee5}.

%\end{document} 
%%%%%%%%%%%%%%%%%%%%%%%%%%%%%%%%%%%%%%%%%%%%%%%%%%%%%%%%%%%%%%%%%%XS 
%%%%%%%%%%%%%%%%%%%%%%%%%%%%%%%%%%%%%%%%%%%%%%%%%%%%%%%%%%%%%%%%%%%%% 

\masubsection{The unitary case for finite matrix size} 
 
When the Hamiltonian matrix \reff{hgoe}, 
$$ H = A + \frac{\lambda}{N} G,$$ 
is such that $G$ belongs to the Gaussian Unitary ensemble (GUE), we  
arrive at an  
analogous expression for the generating function. 
 
In this case, we once again proceed from the expression \reff{js2}, 
where the angular brackets now denote an averaging over the GUE. 
We directly cast this expression into integrals over complex and Grassmannian 
variables by making use of the integral identities \reff{id22} and \reff{id3}. 
This yields 
\be 
J^S = (\det S)^N \int \D\phi_+ \D\phi_- \langle e^{-(\ophi_+, C_+ \phi_+)} 
e^{-(\ophi_-, C_- \phi_-)}\rangle, 
\label{jsu} 
\ee 
where the matrices $C_+$, $C_-$ and $S$ are defined through \reff{c-} 
-- \reff{defs}. 
 
The symmetry of the matrices $G$ with respect to unitary  
transformations allow us to use complex variables. There is no necessity to 
double the dimension of the matrices so as to accomodate real variables, 
as we did in the case of the GOE. Hence the label $p$ which appeared in 
the relations for the GOE, and resulted from this doubling of the dimension, 
do not appear in the corresponding relations for the GOE. All $2n \times 2n$ 
matrices appearing in the case of the GOE are replaced by  
analogous $n\times n$ 
matrices labeled by a single index $\alpha$. 
 
Evaluating the average over the GUE in \reff{jsu}, we obtain an expression  
similar to \reff{jay2}, with the following changes 
\bea 
(\det \cS)^{N/2} &\longrightarrow& (\det S)^N \\  
\Upsilon_- &\longrightarrow& \idty_n  
\label{replace} 
\eea 
We notice that in this case the symmetry in the Grassmannian sector is simply  
the unitary one. This implies that the matrix $Q_-$ is now simply an  
$n\times n$ self-adjoint matrix. The $n \times n$ matrix $Q_+$ has the same  
block structure as before but now $Q_{ii}$, with $i=1,2$, are self-adjoint  
matrices and 
\be 
Q_{12} = Q_{21}^\dagger. 
\ee 
By proceeding exactly as before, we finally arrive at the following 
expression for the generating function: 
\bea 
J^S &=&\frac{1}{d_+\, d_-}\, \int DQ_+ \int DQ_- \,\exp\left(-\frac{1}{2} 
\tr (Q_+ + \frac{i \rho_+ S}{\Lambda})^2 \right)\nonumber\\ 
&& \quad \exp\left(-\frac{1}{2} 
\tr (Q_- -\frac{\rho_-}{\Lambda})^2 \right)\, {\widetilde{K}}(q_+, q_-)  
\label{jay44} 
\eea 
where 
\be 
{\widetilde{K}} = \int D \ochi D \chi e^{-(\ochi, \chi)} P^N. 
\label{hkk} 
\ee 
Here $\ochi, \chi$ are Grassmannian vectors of length $n$, and $P$ is given 
by 
\be 
P:= \int da \, \nu(a)\, \left[\frac{\det X_-}{\det X_+}\right] 
\, \det\left( \idty_{n} +  \frac{\lambda^2}{N^2} \frac{M(X_+)}{X_-}  
\right), 
\label{pee55} 
\ee 
where  
\be 
M_{\alpha_1 \alpha_2} = \sum_{ \gamma=1}^n 
\frac{\chi_{\alpha_1}(\gamma) \,\ochi_{\alpha_2}(\gamma')}{X_+( \gamma)}; 
\ee 
and $X_\pm$ are diagonal $n \times n$ matrices with diagonal elements 
\bea 
x_+(\gamma) &:=& i(e - a) + i \frac{\lambda}{N} q_+(\gamma)\\ 
x_-(\gamma)  &:=& i(e - a) +  \frac{\lambda}{N} q_-(\gamma) 
\label{xx4} 
\eea 
respectively. 
Equivalently, we could also write $P$ in a form analogous to \reff{pee2}, 
\be 
P:= \int da \, \nu(a)\, \det\left( \frac{\lambda}{N} \, q_- 
+ i(e - a)\idty_n \right)  \, \det\left[ i(e - a)\idty_{n} +  
i \frac{\lambda}{N} q_+
+ \frac{\lambda^2}{N^2} \R \right]^{-1}, 
\label{pee24} 
\ee 
with the matrix $\R$ being defined by \reff{arr} as before, 
but the label $\beta$ taking the values $1$ to $n$ only. 
 
The above relations, \reff{jay44} -- \reff{pee55}, for the GUE, are found to 
be very similar in form to the corresponding relations \reff{jay4} --  
\reff{pee5} 
for the GOE. The only difference lies in the fact that the expression 
for $P$ involves {\em{square roots}} of determinants for the GOE 
[see \reff{pee5}], whereas there is no square root appearing in the 
corresponding relation \reff{pee55}, in the case of the GUE.

%------------------------------------------------------------- 
\masubsection{Supersymmetric formulation for the orthogonal and unitary case} 
%%%%%%% begin insertion %%%%%%%%%%%%%%%%%% 
Up to now we have used Grassmannian variables only as auxiliary variables
which are to be finally eliminated by integration.
We can, however, alternatively, cast the expressions 
for the generating function in a supersymmetric form, where Grassmannian 
and ordinary variables are put on the same footing. If we introduce the usual 
parameter $\beta$ taking the value $1$ for the orthogonal case and $2$ for 
the unitary one, then the generating function in both cases can be expressed 
in the following elegant form: 
\be
J^S_\beta = \frac{\int D\qq^\beta \exp\left({-\frac{1}{2} {\hbox{Str}} 
(\qq^\beta - {\rr^\beta}/{\lambda})^2}\right)\, F_\beta (\qq^\beta)^N}
{\int D\qq^\beta \exp\left({-\frac{1}{2} {\hbox{Str}}(\qq^\beta)^2}\right)}
\label{ss}
\ee
where
\be
F_\beta(\qq^\beta) = \int da \, \nu(a) \, {\hbox{Sdet}} 
\left((e -a)\idty_{4n/\beta}
+ \frac{\lambda}{N} \qq^\beta\right)^{-\beta/2},
\label{ss2}
\ee
the ${4n/\beta}\times {4n/\beta}$ supermatrix $\qq^\beta$ 
being given by 
\be
\quad \quad \qq^\beta = 
\left( \begin{array}{lr} Q_+^\beta & Q_{+-}^\beta \\
		  Q_{-+}^\beta & -i Q_-^\beta 
\end{array} \right),
\label{qq}
\ee
and $\rr^\beta$ denoting the diagonal supermatrix 
\be
\quad \quad \rr^\beta = 
\left( \begin{array}{lr} - i \rho_+ {S_\beta} & 0\\
		  0 & -i \rho_-
\end{array} \right).
\ee
Here 
\bea
S_\beta &=& \idty_2 \otimes S \quad \, {\hbox{for}} \quad \beta = 1\\
	&=& S \quad \, {\hbox{for}} \quad \beta = 2.\\
\eea
The measure of the superintegral is
$$ D\qq^\beta := DQ_+^\beta DQ_-^\beta  DQ_{+-}^\beta DQ_{-+}^\beta.$$

Let us show this in the orthogonal case. We start from the expression 
\reff{last} for $\widehat{K}$ and we note that $Z$ [\reff{pee}] can be 
written as  
\be 
(\det \cS)^{-\frac{1}{2}} \int da \, \nu(a) C^{-\frac{1}{2}} 
\ee 
where $C$ can be written as 
\bea
C &=& \det\left[(e - a)\idty_{2n} - i\frac{\lambda}{N} \,Q_-\right]^{-1} 
\,  \nonumber\\
& & \,\, \times \,\, 
\det\left[(e - a)\idty_{2n} - \frac{\lambda}{N} \,Q_+  
- \olambda U^{-1} \,\Bigl((e -a)\idty_{2n} - 
i\frac{\lambda}{N} \,Q_-\Bigr)^{-1}
U \Lambda \right], \nonumber\\
\eea 
where  
\bea 
\olambda(\beta|p\alpha) &=& \frac{1}{\sqrt{2}} \opsi_\alpha(\beta)\\ 
\Lambda(p\alpha| \beta)  &=& \frac{1}{\sqrt{2}} \Psi_\alpha(\beta),
\eea 
for $\alpha = 1 \ldots n$, and $U$ is the matrix which diagonalizes $Q_-$. 
 
In this form we see that $C$ is a superdeterminant: 
\be 
C = {\hbox{Sdet}} \Bigl((e - a)\idty_{4n} + \frac{\lambda}{N} \qq\Bigr), 
\ee 
where 
\be 
\quad \quad \qq =  
\left( \begin{array}{lr} Q_+ & Q_{+-} \\ 
		  Q_{-+} & -i Q_-  
\end{array} \right), 
\label{qq2} 
\ee 
and  
\be Q_{+-} = \olambda U^{-1} \quad \quad Q_{-+} = U \Lambda. 
\ee 
Noting that $ \tr (Q_{+-}\,Q_{-+}) = (\opsi, \Psi)$, we can easily 
see that the exponential term in $J^S$, \reff{jay44}, is indeed of the form 
$$ 
-\frac{1}{2} {\hbox{Str}} (\qq - \frac{\rr}{\lambda})^2. 
$$ 
The unitary case is treated in the same way. 
 
We would like to stress that here the supersymmetric formalism 
gives us only an elegant notation. The formalism itself is only useful 
when we want to mix Grassmannian and real or complex variables. Such 
a mixing arises in the standard case in which the matrix $A=0$. This
case has been treated by Efetov \cite{efe}, by using the 
supersymmetric technique. In the large-$N$ limit a saddle 
point is constructed which mixes up the Grassmannian and ordinary 
variables. The supersymmetric 
technique is also useful in the case where $\lambda = \sqrt{N}$ and  
the matrix $A$ is non-zero, where one can show that the energy level correlations
agree with the predictions of standard Random  
Matrix theory. However, in the model that we are mainly concerned with 
in this paper, namely the one in which $\lambda$ remains fixed for
large $N$, the supersymmetric formulation does not give us any additional
advantage.

\masubsection{The orthogonal and unitary case in the 
limit of infinite matrix size} 
We will now evaluate the generating function in the limit  
$N \longrightarrow \infty$, from which the correlation functions can 
be obtained by taking various derivatives with respect to the
variables $\epsilon_-(\alpha)$. 
It can be easily shown that there is no problem in the interchange of the 
limit $N \longrightarrow \infty$, with the derivation with respect to 
$\epsilon_-$, and the subsequent limit 
$\epsilon_- = \epsilon_+ \longrightarrow 0$. 
 
The result for the generating function both in the orthogonal ($\beta=1$) 
and the unitary case ($\beta=2$) can be put in the following form:
\bea  
J^S &=& \frac{\exp\Bigl(\frac{ip\beta}{2} \tr(\eps_- - \eps_+) S\Bigr)}{d_+ 
d_-} 
\int \, DQ_+ Dq_- D\ochi D \chi \exp\left(-\frac{1}{2}  
\tr\bigl[ Q_+ + \frac{i \rho_+ S}{\lambda} - \frac{\beta \lambda p}{2} \Bigr]^2 
\right) 
\nonumber\\ 
& & \quad \times \, \exp\left(-\frac{1}{2}  
\tr\bigl[ Q_- - \frac{\rho_-}{\lambda} - \frac{i\beta \lambda p}{2} \Bigr]^2 
\right) - \left(\ochi, \chi\right) \nonumber\\ 
& & \quad \times \, \exp \left( \lambda \nu (e) \Bigl[ \frac{\beta A_o}{2} +  
A_1^\beta + A_2^\beta \Bigr] \right).  
\label{genone} 
\eea 
Here $\eps_{\pm}$ are $n \times n$ matrices with diagonal elements 
$\eps_{\pm}(\alpha)$, and 
\be 
p = p(e) = \pp \int da \frac {\nu (a)}{a - e},  
\ee 
with the symbol $\pp$ denoting the principal value of the integral. The matrix
$A_0$ is given by
\be 
A_0 = i \pi \tr (q_+ + i q_-) \sigma(q_+), 
\ee 
where 
$\sigma (q_+)$ is the diagonal matrix whose elements are the signs of the 
imaginary parts of the  eigenvalues $q_+(p\gamma)$ of the matrix $Q_+$. 
\be 
A_1^\beta = \int_{- \infty}^{+\infty} dt \,  
\left\{ \left( \frac{\det(t -iq_-)}{\det (t + q_+)}\right)^{\frac{\beta}{2}} - 
\Bigl[ 1 - \frac{\beta}{2} \tr \frac{q_+ + i q_-}{t + q_+} \Bigr]  
\right\} 
\ee 
and 
\be 
A_2^\beta = \int_{- \infty}^{+\infty} dt \,  
 \left( \frac{\det(t -iq_-)}{\det (t + q_+)}\right)^{\frac{\beta}{2}} 
\, \Bigl[ \det(1 - R^\beta)^\frac{\beta}{2} - 1\Bigr], 
\label{a22}
\ee 
where  
\be 
R_{\alpha \alpha'}^1 = \frac{1}{t - iq_- (\alpha)}\, 
\sum_{p=1}^2\,\sum_{\gamma = 1}^ n \frac{ \ochi_\alpha (p \gamma) \chi_{\alpha'}  
(p \gamma)}{t + q_+(p \gamma)} 
\ee 
and 
\be 
R_{\alpha \alpha'}^2 = \frac{1}{t - iq_- (\alpha)}\, 
\sum_{\gamma = 1}^n \frac{ \ochi_\alpha ( \gamma) \chi_{\alpha'}  
( \gamma)}{t + q_+(\gamma)} 
\ee 
This is the main result of this paper. We have expressed the generating  
function for the correlation functions in terms of a finite set of integrals. 
Hence we have reduced the problem of the computation of the generating
function, in the limit of infinite matrix size, to that of the evaluation 
of a finite set of integrals. This was our main purpose, since, starting
from this explicit expression, we can proceed to evaluate the 
physically relevant correlation functions. 
However, as we shall see, the task of evaluating these integrals is  
non-trivial. Nevertheless, a general conclusion can be drawn from this 
expression by noting that the generating function has the following structure: 
\be 
J^S = \exp\left(\frac{ip\beta}{2} \tr\Bigl[(\eps_- - \eps_+) S\Bigr]\right)\, 
K^S \left(\nu \eps_-, \nu \eps_+, \{ \nu r_\alpha +  
\frac{\lambda^2 \beta \nu \,p}{2}\}, \Lambda\right), 
\ee 
where $\Lambda = \lambda \, \nu (e)$, can be called the  
{\it{renormalized coupling constant}}.

Since the correlation functions can be computed from the generating function  
by using the formula 
\be 
\left(\frac{1}{2\pi}\right)^n \prod_{\alpha= 1}^{n}  
\frac{\partial}{\partial \eps_- (\alpha) } J^{(\idty_n , - \idty_n)}\ 
\biggr|_{\eps_- = \eps_+ = 0} = \rho^{(n)}(r_1, \ldots, r_n), 
\ee 
where  
\be 
J^{(\idty_n , - \idty_n)} = \langle \prod_{\alpha=1}^{n}  
\frac{{\hbox{det}}\,[ 
\epsilon_-^2 (\alpha) + \, (e_\alpha - H)^2]}  
{{\hbox{det}} \,[\epsilon_+^2 (\alpha) + 
\, (e_\alpha - H)^2]} \rangle, 
\ee 
is positive, it follows that  
$$ J^{(\idty_n , - \idty_n)} = |K^{(\idty_n , - \idty_n)}|,$$ 
and therefore $\rho^{(n)}$ has the structure 
\be 
\rho^{(n)}(r_1, \ldots, r_n) = \nu^n \, f^{(n)}_\beta  
(r_1 \nu + a, r_2 \nu + a, \ldots , r_n \nu + a ; \Lambda), 
\label{corr3} 
\ee 
with $ a = \lambda^2 \beta /2$. 
However, since the correlation functions are translation invariant,  
the RHS of \reff{corr3} does not depend on $a$.  
 
We shall prove that the 
density of states $\rho (e)$ is equal to $\nu (e)$. 
We can therefore conclude that, {\it{on the scale of energy where 
the mean level spacing is equal to unity}}, the correlation 
functions are {\it{universal}}, i.e., they depend only on $\beta$ and 
$\Lambda$. More precisely, 
\be 
\left[\frac{1}{\rho (e)}\right]^n \, \rho^{(n)} \Bigl(  
\frac{r_1}{\rho (e)}, \ldots, \frac{r_n}{\rho (e)} \Bigr) 
= f^{(n)}_\beta \Bigl( r_1, r_2, \ldots, r_n ; \Lambda\Bigr). 
\ee  
Let us now derive eqn. \reff{genone}. We decompose $P$, as given by
\be
P = \int da \, \nu (a) \, 
\left(\frac{\det X_-}{\det X_+}\right)^{\frac{\beta}{2}}
\, \left( \det \Bigl[\idty + \frac{\lambda^2 M(X_+)}{ N^2 X_-} \Bigr]   
\right)^{\frac{\beta}{2}},
\ee
into three terms, i.e., $P = P_0 + P_1 + P_2$, where
\be
P_0 := 1 + \frac{\beta}{2} \int da \, \nu(a) \tr \frac{X_- -  X_+}{X_+},
\label{p0}
\ee
\be
P_1 :=  \int da \, \nu(a) \left\{\left(\frac{\det X_-}{\det X_+}\right)^{\beta/2}
\, - \left[ 1 + \frac{\beta}{2} \tr \frac{X_- -  X_+}{X_+} \right] \right\}
\label{p1}
\ee
and
\be
P_2 :=  \int da \, \nu(a) \left(\frac{\det X_-}{\det X_+}\right)^{\beta/2}
\, \left\{ \left[ \det (1 + \frac{\lambda^2 M(X_+)}{N^2 X_-})\right]^{\frac{\beta}{2}} 
- 1\right\}.
\label{p2}
\ee
Let us first evaluate $P_0$ in the large-$N$ limit.
\be
\int da \, \nu(a) \tr \frac{X_- -  X_+}{X_+} = - \frac{\lambda}{N} 
\int da \, \nu(a) \tr \left[(q_+ + i q_-)
\,  \Bigl({ (e -a)\idty_{2n} + \frac{\lambda}{N} q_+}\Bigr)^{-1}\right].
\label{abb}
\ee
since $\Imm q_+ (\gamma) \ne 0$, the integral on the RHS of \reff{abb}
tends to the expression
$$ - \Alpha(e)  - i\pi \nu (e) \sigma_+ (p \gamma)$$
as $N \longrightarrow \infty$, where $\sigma_+ (p \gamma)$ denotes 
the sign of the imaginary part of the
eigenvalue $q_+(p \gamma)$.
\medskip

\noindent
Hence, for large $N$, 
\be
P_0 = 1 + \frac{\lambda}{N} \frac{\beta}{2} \left[ p \tr (q_+ + i q_-)
+ i \pi \nu \tr (q_+ + i q_-) \sigma (q_+) \right],
\ee
where $\sigma (q_+)$ is the diagonal matrix with elements $\sigma_+ (p\gamma)$.
In the second term, $P_1$, we make the change of variables $e -a = 
{t \lambda }/{N}$, so that it reads
\be
P_1 = \frac{\lambda}{N} \int_{- \infty}^{+\infty} dt \, \nu \Bigl( e - 
{t \lambda}/{N}\Bigr) \left\{ \left( \frac{\det(t -iq_-)}{\det(t + q_+)} 
\right)^{\frac{\beta}{2}} - 
\left[ 1 - \frac{\beta}{2} \tr\frac{q_+ + iq_-}{t + q_+} \right] \right\}.
\ee
The term in the paranthesis is bounded in $t$ and decays like $ 1/t^2$ when 
$t$ is large, since $\Imm q_+ (\gamma) \ne 0$. Hence, 
we can use the dominated convergence theorem \cite{reed} to show that if
$${{\hbox{sup}}\atop{t}} \, \nu (t)  < \infty,$$
and $\nu (t)$ is continuous, then for large $N$,
\be
P_1 = \frac{\lambda \nu (e)}{N} \, A_1^\beta.
\ee
The term $P_2$ is treated in exactly the same way as $P_1$, so that
asymptotically,
\be
P_2 =\frac{\lambda \nu (e)}{N} \, A_2^\beta.
\ee 
Finally,
\be
{\lim\atop{N \rightarrow \infty}} P^N = \exp\left(
\frac{\lambda \beta p}{2} \, \tr (Q_+ + iQ_-) + \lambda \nu \Bigl[
\frac{\beta}{2} A_0 + A_1^\beta + A_2^\beta\Bigr]\right).
\label{expp}
\ee
Here and henceforth, we write $\nu$ for $\nu(e)$. 
The expression given by eqn. \reff{genone} for $J^S$ is obtained by
completing the square in $Q_+$ and $Q_-$.
\smallskip

\noindent

\masubsection{The density of states and the average of the product of traces of
advanced Green's functions}
\label{average}
The only computation which is easy in the general case, is that of the 
generating function for traces of advanced Green's functions. This corresponds
to the choice $s_\alpha =1 $ for all $\alpha \in \{1 \ldots n \}$.

Let $q_{\pm}(j)$ denote the eigenvalues of the matrices $Q_\pm$.
For the above-mentioned choice of the matrix $S$, 
we know that $q_+ (j) = q^\prime_+ (j) - i \delta$,
where $q^\prime_+ (j)$ is real and $\delta$ is positive. Since
$q_- (j)$ is also real (and doubly degenerate in the $\beta = 1$ case)
we see that the integrands in the expressions for $A_1^\beta$ and 
$A_2^\beta$ are analytic in the variable $t$ in the lower half-plane, and decay
like $1/t^2$. We can therefore apply Cauchy's theorem to simply conclude
that $A_1^\beta = A_2^\beta = 0$. In contrast, if we computed these 
quantities for the case of a mixed product of 
advanced and retarded Green's functions
$( s_\alpha =1, \, \alpha = 1 \ldots p, \, s_\alpha = -1, \, \alpha = p+1, \ldots n)$, there would be $n-p$ singularities in the lower half-plane,
and, therefore, $A_1^\beta$ and $A_2^\beta$ would be non-zero.

Hence, it follows easily from \reff{genone} that the generating function
factorizes as follows:
\be
J^{\idty_n} = \exp \left( i (p + i \pi \nu) \sum_{\alpha=1}^n[\eps_-(\alpha)
- \eps_+ (\alpha) ] \right) J_+ \, J_-,
\ee
where
\be
J_+ = \frac{1}{d_+} \int D Q_+ \, \exp \left (- \frac{1}{2}
\tr \Bigl[ Q_+ + \frac{i \rho_+ S}{\lambda} - \frac{\beta \lambda}{2} 
(p + i \pi \nu) \Bigr]^2\right ),
\ee
and
\be
J_- = \frac{1}{d_-} \int D Q_- \, \exp \left (- \frac{1}{2}
\tr \Bigl[ Q_- - \frac{ \rho_-}{\lambda} - \frac{i\beta \lambda}{2} 
(p + i \pi \nu) \Bigr]^2 \right ).
\ee
Hence, we see from the definitions of $d_\pm$ [\reff{dpm}] that 
$J_\pm = 1$ and
\be
J^{\idty_n} = \exp \left( i (p + i \pi \nu) \sum_{\alpha=1}^n
[\eps_-(\alpha)
- \eps_+ (\alpha) ] \right).
\ee
This implies that 
\be
{\lim\atop{N \rightarrow \infty}} \langle \prod_{\alpha = 1}^{n}
\frac{1}{N} \, \tr G_{e_\alpha}^+ \rangle = [ - p - i\pi \nu]^n,
\ee
which in turn shows that average of a product of the traces of advanced 
(or retarded) Green's functions factorize. In particular, we see that the 
density of states, $\rho (e)$, is simply
given by
\be
\rho (e) = \nu (e).
\ee

%%%%%%%%%%%%%%%%%%%%%%%%%%%%%%%%%%%%%%%%%%%%%%%%%%%%%%%%%%%%%%%%%%%%%%%%%%
\masection{The perturbed Wigner-Dyson ensembles}
%%%%%%%%%%%%%%%%%%%%%%%%%%%%%%%%%%%%%%%%%%%%%%%%%%%%%%%%%%%%%%%%%%%%%%%%%%
We can also look at the situation considered by Brezin and Hikami in the 
unitary case, that is, the situation in which the coupling constant
$\lambda$ is of the order of $\sqrt{N}$. Up to a trivial rescaling, we can
simply take $\lambda = \sqrt{N}$. We will in this case start from the 
supersymmetric formula \reff{ss}, which we repeat here for convenience.
\be
J^S_\beta = \frac{\int D\qq^\beta \exp\left({-\frac{1}{2} {\hbox{Str}} 
(\qq^\beta - {\rr^\beta}/{\lambda})^2}\right)\, F_\beta (\qq^\beta)^N}
{\int D\qq^\beta \exp\left({-\frac{1}{2} {\hbox{Str}}(\qq^\beta)^2}\right)}
\ee
where
\be
F_\beta(\qq^\beta) = \int da \, \nu(a) \, {\hbox{Sdet}} 
\left((e -a)\idty_{4n/\beta}
+ \frac{\lambda}{N} \qq^\beta\right)^{-\beta/2},
\ee
the ${4n/\beta}\times {4n/\beta}$ supermatrix $\qq^\beta$ 
being given by 
\be
\quad \quad \qq^\beta = 
\left( \begin{array}{lr} Q_+^\beta & Q_{+-}^\beta \\
		  Q_{-+}^\beta & -i Q_-^\beta 
\end{array} \right),
\ee
and $\rr^\beta$ denoting the diagonal supermatrix 
\be
\quad \quad \rr^\beta = 
\left( \begin{array}{lr} - i \rho_+ {S_\beta} & 0\\
		  0 & -i \rho_-
\end{array} \right).
\ee
The measure of the superintegral is
$$ D\qq^\beta := DQ_+^\beta DQ_-^\beta  DQ_{+-}^\beta DQ_{-+}^\beta.$$
%**********
Following Efetov, \cite{efe}, let us diagonalize the supermatrix $\qq$, i.e.,
we define 
\be
\qq = V q V^{-1},
\ee
where $q$ is a diagonal supermatrix. Here and henceforth, 
we suppress the superscript
$\beta$, unless explicitly required. The measure $DQ$ hence factorizes 
as follows
\be
DQ = d \mu (\nu) \, m(q) \, dq.
\ee
The generating function will therefore be of the form
\be
J_\beta = \frac{1}{d} \int d\mu (\nu) m(q) dq \, 
\exp \left(- \frac{1}{2} \str ( q - \frac{V^{-1} p V}{\sqrt{N}}\right )^2
\, \langle \sdet [ (e-a)\idty_{m} + \frac{q}{\sqrt{N}}] ^{-\frac{\beta}{2}} 
\rangle^N,
\ee
where the brackets $\langle \cdot \rangle$ denotes an average with respect to 
the measure $\nu (a)$ on $a$ and $m = 4n/\beta$.

Keeping $V$ fixed, we look for a saddle point $\qh$, of order $\sqrt{N}$, 
of the integral over $q$. We take
\be
\qh_+ = \qh_- = \sqrt{N} \, (b \idty_{m/2} - i c S_\beta), 
\ee
where $c$ is positive and $b$ is real.
Since $\str \, \qh^2 =0$ and
$\sdet [(e -a)\idty_{m} + \qh] = 1$, we see that the condition for 
$\qh$ to be a saddle point is
\be
\frac{\qh}{\sqrt{N}} = - \frac{\beta}{2} \langle r \rangle,
\label{up}
\ee
where 
\be
r = \left( (e - a)\idty_{m} + \frac{\qh}{\sqrt{N}}\right)^{-1}.
\ee
Note that for a function $f(a)$, the symbol $\langle f(a) \rangle$
denotes the average
\be
\langle f(a) \rangle = \int da \, \nu (a)\, f(a).
\ee
The condition \reff{up} is satisfied if $z = b - ic$ is a solution of the equation
(first obtained by Pastur \cite{pastur}):
\be
z + \frac{\beta}{2} \langle (e-a + z)^{-1} \rangle = 0.
\ee
Taking into account the contribution of the fluctuations around this
saddle-point, we see that, if ${\cS}$ 
is the supermatrix
$$\cS = \idty_2 \otimes S,$$
then, to order $1$ in $1/(\sqrt{N})$, we have 
\be
J = \frac{1}{d} \, \exp\Bigl( b\,  \str (p)\Bigr) \,
\int d \mu (\nu) \, m (\qh ) \, F \, \exp \left( - i c \str ( V {\cS} V^{-1}
p ) \right),
\ee
where
\be
F = \int d (\delta q) \exp (g),
\ee
with 
\be
g = - \frac{1}{2} \str (\delta q )^2 + \frac{\beta}{4} 
\str \langle (r \, \delta q )^2 \rangle  + \,
\frac{\beta^2}{8}  \left[ \langle \Bigl( \str (r \delta q)\Bigr)^2 \rangle - 
\langle \str (r \delta q) \rangle ^2 \right].
\ee
However, we see that $F$ and $m (\qh)$ depend only on $b$ and $c$, and hence
the generating function $J^S$ has the structure
\be
J^S = \exp \left( ib \tr \bigl[(\eps_- - \eps_+) S\bigr]\right) \, C(b, c)
\, K( c\eps_+, c\eps_- , cr ),
\ee
$K$ being the same function as the one in the case where the matrix $A$
is zero,  i.e.,
the standard case of Random Matrix Theory. It is easy to see that
$c = \pi \rho (e)$, where $\rho (e)$ is the density of states. Therefore,
proceeding as before, we conclude that,
on the scale of energy in which the mean level spacing is unity, 
 the {\it{correlation functions}} are the {
\it{same
as those of Random Matrix Theory}}. More precisely,
\be
\left( \frac{1}{\rho (e)} \right)^n \, \rho^{(n)} \left(
\frac{r_1}{\rho (e)}, \frac{r_2}{\rho (e)}, \ldots, \frac{r_n}{\rho (e)}
\right) = \rho_\beta^{(n)} (r_1, \ldots, r_n),
\ee
$\rho_\beta^{(n)}$ being $n$-point correlation function of the
standard Random Matrix Theory ensemble, characterised only by $\beta$. 

In the unitary case ($\beta = 1$), this conclusion agrees with that
of Brezin and Hikami, \cite{brezin}, who considered an ensemble with 
a fixed matrix $A$. Even though their result appears formally to be
valid for any set of values of $a_j$ ($a_j$ being the diagonal elements
of $A$), it is clear that at best it is true with probability one
with respect to some probability distribution on the $a_j$'s. In contrast,
our result concerns correlation functions which are averaged over the 
$a_j$'s.

In the case $\alpha > 1$, we take $\lambda = 1/(N^{\alpha -1})$ in 
the above expressions, and it is easily seen that when $N$ is large, the
dominant term in the generating function is the generating function 
for the matrix $A$. In other words, the statistics is Poissonian in
this case.

In order to proceed further, we will now look at the simplest case, that
of the two-point correlation functions.

%%%%%%%%%%%%%%%%%%%%%%%%%%%%%%%%%%%%%%%%%%%%%%%%%%%%%%%%%%%%%%%%%%%%%%%%%%
\masection{Unitary case: The two-point generating function}
%%%%%%%%%%%%%%%%%%%%%%%%%%%%%%%%%%%%%%%%%%%%%%%%%%%%%%%%%%%%%%%%%%%%%%%%%%
There is one major simplification in the unitary case. The integral defining
$A_2$ [\reff{a22}] can be explicitly evaluated and gives a 
{\em{meromorphic}} function
of the eigenvalues $\{q_{-j}\}$ and $\{q_{+j}\}$ of the 
matrices $Q_-$ and $Q_+$, respectively. We will, however, only consider the case
of the two-point generating function ($n=2$), with $s_1=+1$ and $s_2= -1$.
The result of the integration over $t$ can be expressed as:
\bea
A_2 &=& \frac{\pi i}{2 w^2}\bigl[ (x_2 + iy_1) C_{22}^1 
+ (x_2 + iy_2) C_{11}^1 -(x_1 + iy_1) C_{22}^2  -(x_1 + iy_2) C_{11}^2 \bigr] 
\nonumber\\
& & \quad - \frac{\pi i}{4w^3}
\bigl[ \det C^1 + \det C^2 - Z^2\bigr],
\eea
where 
\be
x_j = q_{+j} \quad;\quad y_j = q_{-j} \quad 
\ee
and
\be
w = \frac{x_1 - x_2}{2} 
\ee
We choose $\Imm x_1 <0 $ and  $\Imm x_2 > 0 $ so that $\Imm w <0$;
The matrix $C^\alpha$, with $\alpha\in \{1,2\}$, has for elements 
\be
C^\alpha_{ij} = \ochi^\alpha_i \chi^\alpha_j,
\ee
while
\be
Z = C^1_{11} C^2_{22} - C^1_{12} C^2_{21} + C^2_{11} C^1_{22} - C^2_{12}
C^1_{21}.
\ee
We first integrate over the Grassmannian variables 
$\{\chi^\alpha_j, \ochi^\alpha_j\}$ so that if we define
\be
I := \int D\ochi D \chi \exp \bigl( - \sum_{\alpha =1}^2  
\tr C^\alpha + \lambda \nu A_2\bigr),
\ee
we find that 
\be
I = F(y_1)F(y_2) g(y_1) g(y_2) + 3 d^2 + d [F(y_1) -g(y_1)]\,
[F(y_2) -g(y_2)],
\ee
with 
\bea
F(y)&=&  1 - \frac{\pi i \lambda \nu}{w^2} (x_2 + iy) \\
g(y) &=& 1 + \frac{\pi i \lambda \nu}{w^2} (x_1 + iy) \\
\eea
and
\be
d =- \frac{\pi i \lambda \nu}{2 w^3}.
\ee
Grouping these results we can express the generating function as:
\bea
J^{+-} &=& \frac{1}{d_+d_-} \int dQ_+ \int dQ_- 
\exp\Bigl[ - \frac{1}{2} \tr (Q_+ + \frac{i\rho_+ S}{\lambda})^2
-  \frac{1}{2} \tr (Q_- - \frac{\rho_-}{\lambda})^2 \Bigr] \nonumber\\
& &\times \, I \, \exp\Bigl( - \lambda \alpha [x_1 + x_2 + i(y_1 + y_2)]
+ i \pi \lambda \nu [ - 2w + i(y_2 - y_1)]\Bigr)\nonumber\\
& & \times \exp\Bigl( - \frac{\pi i \lambda \nu}{w} 
(x_1 + iy_1) (x_2 + iy_2) \Bigr)
\eea 
We now integrate over $Q_-$. The integral has the structure
\be
A = \int dQ_- F(y_1, y_2) \exp \Bigl( -\frac{1}{2} \tr Q^2_{-}
+ \frac{1}{\lambda} \tr (\rho_- Q_-)\Bigr),
\label{aa}
\ee
where $F(y_1, y_2)$ is some symmetric function of $y_1$ and $y_2$. If we
define
\bea
z_1 &:=& \frac{y_1 - y_2}{2} \\
z_2 &:=& \frac{y_1 + y_2}{2}, 
\eea
we can write
$$
Q_- = \left( \begin{array}{lr} z_2 + z_1 \cos \phi & z_1 \sin \phi e^{i\psi}
 \\
		 z_1 \sin \phi e^{-i\psi}   &  z_2 - z_1 \cos \phi 
\end{array} \right)
$$
with $\phi \in [0, \pi]$ and $\psi \in [0, 2\pi]$. In terms of the variables
$z_1$ and $z_2$, we find that 
\bea
A &=& \frac{2\pi \lambda}{\tr (\rho_- s)} \sum_{\sigma \in \{-1, +1\}} \sigma
\int_{- \infty}^{+ \infty} dz_1 \int_{- \infty}^{+ \infty} dz_2 \, z_1\,
F(z_1 + z_2, z_2 - z_1)\nonumber\\
&& \times \, \exp\Bigl(-[z_1^2 + z_2^2] + \frac{z_2}{\lambda} \tr \rho_- +
\frac{\sigma z_1}{\lambda} \tr(\rho_- S)\Bigr)
\eea
and the generating function is given by the expression
\bea
J^{+-} &=& \frac{2 \pi \lambda}{d_+ d_- \tr (\rho_- S)}
\exp\Bigl( \frac{1}{2\lambda^2}[\tr(\rho_+ S)^2 - \tr \rho_-^2]\Bigr)
\nonumber\\
&& \times \sum_\sigma \sigma \int dQ_+ \exp \bigl[ - \frac{1}{2} \tr Q_+^2 -
\frac{i}{\lambda} \tr Q_+ \rho_+ S\bigr] \, G_\sigma (z_3 , w),
\eea
where
\be
z_3 = \frac{x_1 + x_2}{2}
\ee
and
\bea
G_\sigma (z_3, w) &=& \int dz_1 \int dz_2 z_1\, I\,
\exp\bigl[ - z_1^2 ( 1 + \frac{i\pi \lambda \nu}{w} )
- z_2^2 ( 1 - \frac{i\pi \lambda \nu}{w} ) \bigr] \nonumber\\
& &  \exp\bigl(- \pi i \lambda \nu w\bigr) \, \exp \bigl( \frac{2 \pi \lambda \nu}{w}
z_2 z_1 - 2 \lambda \alpha [z_3 + iz_2]\bigr).
\eea 
This last integral converges if $|\Imm w| > \pi \lambda \nu$, a condition
that we can impose by choosing the free parameter $\delta$ appearing in 
$Q_+$.

Now the matrix $Q_+$ has the form
$$
Q_+ = \left( \begin{array}{lr} z_3 + q - i \sqrt{\delta^2 +h}
& \sqrt{h}e^{i\theta}
 \\
\sqrt{h}e^{-i\theta}		    &
z_3 - q + i \sqrt{\delta^2 +h}
\end{array} \right)
$$
where $\theta \in [0, 2 \pi]$, $h \ge 0$, $q \in \Rl$ and 
$dQ_+ = 2dz_3\, dq\, dh\,d\theta$.
Since
\be
\frac{1}{2} \tr Q_+^2 = w^2 + z_3^2 \quad \quad {\hbox{with}} \quad 
w = \sqrt{(q - i \sqrt{\delta^2 +h})^2 + h},
\ee
we have that
\be
\tr (Q_+ \rho_+ h) = z_3 \tr (\rho_+ h) + (q - i \sqrt{\delta^2 + h})\, 
\tr \rho_+.
\ee
Hence, we need to compute an integral of the form
\be
B = \int_{-\infty}^{+\infty} dq \int_{0}^{\infty} dh\, 
\exp \Bigl( - w^2 - i \tr \rho_+ (q - i \sqrt{\delta^2 + h}) \Bigr)\,  
G_\sigma (z_3, w).
\ee

It can be shown that since $ G_\sigma (z_3, w)$ is analytic and bounded in 
the domain {$\Imm w < - \pi \lambda \nu$}, we can change the integral 
over $q$ into an integral over $w$ 
along the path $C:= \Imm w = - b$, with $b > \pi \lambda \nu$, by choosing 
$\delta > \pi \lambda \nu$,
so that the integral over $h$ can be evaluated. Finally, we obtain
\be
B = \frac{2i\lambda}{\tr \rho_+} \int_{C} dw \, w\, \exp\Bigl( -[w^2
+ \frac{iw}{\lambda} \, \tr \rho_+]\Bigr) \,  G_\sigma (z_3, w).
\ee
Hence,
\be
J^{+-} = \frac{8 \pi^2 i \lambda^2 \exp \Bigl( \frac{1}{2\lambda^2} \tr (\rho_+^2 - 
\rho_-^2)\Bigr)}{(\tr \rho_+) (\tr \rho_- s) d_+ d_-} \,
\sum_\sigma \sigma \int_{C} dw \, w\, \exp\Bigl( -[w^2
+ i \pi \lambda \nu w + \frac{iw}{\lambda} \, \tr \rho_+]\Bigr) \, K_\sigma (w),
\ee
where
\bea
K_\sigma (w) &=& \int dz_1 dz_2 dz_3 \, z_1 I \exp \Bigl( - \sum_j z_j^2 -
\frac{\pi i \lambda\nu}{w} [ (z_3 + i z_2)^2 + z_1^2 ] - 2 \lambda \alpha [z_3 + i z_2]
\Bigr)\nonumber\\
& & \quad \quad \times \quad \exp \Bigl( \frac{\sigma}{\lambda} z_1 \tr (\rho_- s) 
+ \frac{z_2}{\lambda} \tr (\rho_-) - \frac{i z_3}{\lambda} \tr (\rho_+ s) \Bigr),
\eea
and
\be
I = A_1 + (z_3 + i z_2)^2 A_2 + z_1^2 A_3 + [(z_3 + iz_2)^2 + z_1^2]^2 \, A_4,
\ee
with
\bea
A_1 &=& \Bigl( 1 + \frac{i \Lambda }{2w} \Bigr)^4 - \frac{3 \Lambda^2}{4 w^6}\nonumber\\
A_2 &=& - \frac{\Lambda^2}{2 w^4}\,\Bigl[ - \frac{i \Lambda }{2w^3} - 
\bigl( 1 + \frac{i \Lambda }{2w}\bigr) ^2 \Bigr] \nonumber\\
A_3 &=& - \frac{\Lambda^2}{2 w^4}\,\Bigl[ - \frac{i \Lambda }{2w^3} + 
\bigl( 1 + \frac{i \Lambda }{2w}\bigr) ^2 \Bigr] \nonumber\\
A_4 &=& \frac{1}{16}\frac{\Lambda^4 }{w^8},
\eea
where
\be
\Lambda = \pi \lambda \nu.
\label{lambda}
\ee
We now define 
\bea
M &=& \int dz_1 dz_2 dz_3 \, \exp\Bigl[ - \sum_{j=1}^3 z_j^2 - \frac{c \Lambda}{w} [
(z_3 + iz_2)^2 + z_1^2\Bigr] \nonumber\\ 
& & \quad \times \exp\Bigl[\sigma t z_1 - y (z_3 + i z_2) - u (z_3 - i z_2)\Bigr],
\eea
with
\bea
u &=& \frac{1}{2i\lambda} \, \tr (\rho_- - \rho_+ s)\\
y &=& 2 \lambda \alpha - \frac{1}{2i\lambda}\Bigl[\tr \rho_- + \tr (\rho_+s) \Bigr]\\
t &=& \frac{1}{\lambda} \, \tr (\rho_- s),
\eea
so that
\be
\sigma K_\sigma = \frac{\partial}{\partial t} 
\left[ A_1 + A_2 \, \frac{\partial^2}{\partial y^2} + A_3 \frac{\partial^2}{\partial t^2}
+ A_4 \Bigl( \frac{\partial^2}{\partial y^2} + \frac{\partial^2}{\partial t^2} \Bigr)^2 \right]\, M,
\ee
and the Gaussian integral $M$ is given by
\be
M = \frac{\pi^{3/2}}{(1 + \frac{i \Lambda}{w})^{\frac{1}{2}} } \times  \exp\left[
\frac{t^2}{4(1 + \frac{i \Lambda}{w})} + u y - u^2 \frac{i \Lambda}{w} \right].
\ee
Since $d_\pm = 2 \pi^2$, we can express the generating function as
\be
J^{+-} = \frac{4i}{\pi v t} \, \exp\left[ uy + \frac{1}{2\lambda^2} \tr (\rho_+^2 - \rho_-^2)
\right] \, \partial_t R,
\label{tsix}
\ee
where
\be
R = \langle w A_1 \rangle + u^2 \langle w [A_2 - A_3] \rangle + 
\widehat{L}  \langle w A_3 \rangle + \widehat{L}^2  \langle w A_4 \rangle;
\ee
If $F(w)$ is a function of $w$, we define
\bea
 \langle F(w) \rangle &=& \int_C dw F(w) \exp\Bigl[ - w^2 - i w(v + \Lambda) - 
\frac{i \Lambda u^2}{w}\Bigr]\nonumber\\
& & \quad \quad \times \, \int_{- \infty}^{+ \infty}  dz \, \exp\Bigl[ - z^2 ( 1 + \frac{i \Lambda}{w} ) - tz\Bigr],
\label{two8}
\eea
and $\widehat{L}$ is the operator
\be
\widehat{L} = u^2 +  \frac{\partial^2}{\partial t^2}.
\ee
We can express $R$ as a linear combination of the functions
\be
B_n = \langle w^{-n} \rangle,
\label{three0}
\ee
if we note that 
\be
\widehat{L}B_n = - \frac{i}{\Lambda} (n-2) B_{n-1} -  \frac{2i}{\Lambda} B_{n-3}
+ (v + \Lambda) B_{n-2},
\ee
so that
\be
R = \sum_{j=1}^3 \alpha_j B_j,
\ee
with
\bea
\alpha_{-1}&=& =1 , \,\, \alpha_0 = 2i\Lambda , \,\,  \alpha_1 = - \frac{13}{4} \Lambda^2 -
\frac{v \Lambda}{2}\nonumber\\
 \alpha_2 &=& - \frac{3}{2}i \Lambda^3 +\frac{i \Lambda}{2} - \frac{3}{4} v \Lambda^2, \, \, 
\alpha_3 =\frac{\Lambda^4}{4} - \frac{7}{8} \Lambda^2 + \frac{v \Lambda^3}{4} 
+ \frac{v^2}{16} \Lambda^2 + u^2 \Lambda^2\nonumber\\
\alpha_4 &=& i u^2 \Lambda^3 - \frac{3}{8} i \Lambda^3, \, \, 
\alpha_5 = - \frac{u^2}{4} \Lambda^4.
\eea
One can also use the recursion formula
\be
\partial_t B_n = \frac{t}{2} B_n - i \Lambda \partial_t B_{n+1}
\label{recurse}
\ee
to express the generating function in terms of the $B_n$, for $n \in \{ -1, 4\}$ and
$\partial_t B_5$.
 
One can also show that $B_n$ can be represented as 
an integral over the modified Bessel functions.
However, we shall not use this representation now, since we do not need to compute the
generating function itself, but only its derivatives.

\masubsection{Unitary case: The two-point correlation function}
In order to compute the correlation function, we start from the expression \reff{tsix}
for the generating function and note that when $\eps_+ = \eps_-$, $u = 0$. Hence, since
$J^{+-} (\eps_+ = \eps_-) = 1$, we have that 
\be
\partial_t R\biggr|_{u=0} = \frac{v t \pi}{4i}.
\ee
Moreover,
\bea
\frac{\partial J^{+-}}{\partial \eps_+ (1)}\biggr|_{\eps_- = \eps_+} &=& - \pi \nu + i \alpha,
\nonumber\\
\frac{\partial J^{+-}}{\partial \eps_+ (2)}\biggr|_{\eps_- = \eps_+} &=& - \pi \nu - i \alpha,
\eea
since these derivatives give the average value of
$$ \langle - \frac{1}{N} \tr \Bigl( \frac{\eps_+(j)}{N} + i (e_j -H)\Bigr)^{-1}\rangle$$
in the limit $N \longrightarrow \infty$. Using these relations, one can show that
\be
\frac{\partial^2 J^{+-}}{\partial \eps_+ (1)\partial \eps_+ (2)}\biggr|_{\eps_- = \eps_+=0}
= \alpha^2 + \frac{1}{\lambda^2} \Bigl[ \frac{3}{2} - \frac{v^2}{4} - iv + \frac{2 \Lambda}{v} 
+ M
\Bigr],
\ee
with
\be
M = \frac{4i}{\pi v^2} \Bigl[ \partial_t\Bigl( \frac{1}{2} \frac{\partial^2 R}{\partial u^2}
+  \frac{\partial^2 R}{\partial v^2}\Bigr)\Bigr]\biggr|_{u=0, v=t= \frac{i}{\lambda}(r_1 - r_2)}.
\ee
Since the two-point correlation function $\rho_2(r_1, r_2)$ is given by
\be
2 \pi^2 \rho_2(r_1, r_2) = \Ree \left[
\frac{\partial^2 }{\partial \eps_+ (1)\partial \eps_+ (2)} (J^{+-} + J^{++}) \right]
\biggr|_{\eps_- = \eps_+=0},
\ee
we see that the unfolded cluster function 
\be
Y(r) = \frac{1}{\nu^2} \left[\nu^2 -\rho_2(\frac{r_1}{\nu},\frac{r_2}{\nu})  \right],
\,\, {\hbox{with}} \,\, r = r_1 - r_2,
\ee
is given by
\be
Y(r) = \frac{1}{2 \Lambda^2}\, \Ree \left( \frac{3}{2} - \frac{v^2}{4} - \Lambda v 
- \Lambda^2 +M \right),
\label{seven}
\ee
where $v = \pi i r/ \Lambda$.

It remains, therefore, to compute $M$. This is a lengthy computation, which is simplified by 
making use of the recursion formula \reff{recurse} and the following relations:
\bea
\partial_{u^2} B_n &=& - i \Lambda B_{n+1} \nonumber\\ 
\partial_{v} B_n &=& - i  B_{n-1}.
\eea
In this way we get 
\be
M= \frac{2i}{\pi v} \sum_{j=-3}^{4} \beta_j B_j^0 + \frac{4i}{\pi v^2} \beta_5 
(\partial_t B_5)^0.
\ee 
The $\beta_j$ are some polynomials of second degree in $v$, and the superscript $0$ indicates
that the quantities are computed with $u=0$ and $v=t={i\pi r}/{\Lambda}$.

It remains to compute the $B_n$. From eqns. \reff{two8} and \reff{three0}, we see that
if we made the change of variables $z = s - iw$, when $u=0$ and $v=t$,
then we could write $B_n$ in the following form
\be
B_n = \int_C dw\, w^{-n} \, \int_{-\infty}^{+\infty} ds \, 
\exp\left( - s^2 -s(v + 2 \Lambda) - i \Lambda s^2 + 2 i w s\right).
\ee
When $n \ge 1$, we can interchange the two integrals and replace the $w$ integral by a contour
integral around the origin when $s \ge 0$ (because $\Imm w < - \Lambda$), whereas,
if $s < 0$, the $w$-integral vanishes. In this way one finds that, when $n \ge 1$
\be
B_n = 2 \pi i \, \int_0^\infty ds \, f(s) (2is)^{n-1} \,
 F_{n-1} (2 \Lambda s^3)
\ee
where $f(s) = \exp\Bigl( - s^2 - s(v + 2\Lambda) \Bigr)$ and
\be
F_n (x) = \sum_{j=0}^\infty \frac{x^j}{j! (n+j)!} = x^{-\frac{n}{2}} \, I_n (2 \sqrt{x}),
\ee
$I_n$ being the modified Bessel function. In order to get this form, we have simply
expanded $\exp ( - i \Lambda s^2/w)$ in powers of $w^{-1}$ in the $w$ integral.

When $n \le 0$,  we have to do the interchange of the $s$-integral and the $w$-integral
more carefully, and one finds that
\be
B_n = B_n^\prime + B_n^{\prime\prime},
\ee
where
\be
B_n^\prime = 2 \pi i \, \int_0^\infty ds \, f(s) (- i \Lambda s^2)^{n+1} \, 
F_{n+1} (2 \Lambda s^3)
\ee
and $ B_n^{\prime\prime}$ is some polynomial of degree $n$ in $v$. Similarly, when $n \ge 1$,
we have that
\be
\partial_t B_n = i B_{n-1} -R_n
\ee 
where
\be
R_n = \pi \int_{0}^\infty ds \, f(s) (2is)^n \, F_{n-1} (2 \Lambda s^3).
\ee
We have now at our disposal all the quantities appearing in $M$, and, 
therefore, from \reff{seven}, the cluster function $Y$. The final result
for $Y$ is given by the following expression;
\bea
Y &=& \int_0^\infty ds \exp\left(-s^2-2\Lambda s\right)\,\Bigl[\sum_{j=0}^4 \alpha_j 
F_j (2 \Lambda s^3)\Bigr] \, \cos \left(\frac{\pi r s}{\Lambda}\right)
\nonumber\\
& & \quad + \,\int_0^\infty ds \exp\left(-s^2-2\Lambda s\right)\,
\Bigl[\sum_{j=0}^4 \beta_j 
F_j (2 \Lambda s^3)\Bigr] \, \sin \left(\frac{\pi r s}{\Lambda}\right)
\eea
where
\bea
\alpha_0 &=& - \frac{(s + \Lambda)}{4} \nonumber\\
\alpha_1 &=& \frac{s^2\,\Lambda}{4} \nonumber\\
\alpha_2 &=& s^4 \Lambda - 3 \Lambda s^2\nonumber\\
\alpha_3 &=& s^3\Lambda^2 \left[2 \Lambda^2 - 3 - 
\frac{8 \Lambda^2}{\pi^2 r^2} \right] \nonumber\\
\alpha_4 &=& s^5\Lambda^4 \left[\frac{16 \Lambda^2}{\pi^2 r^2} - 2 
- 4 \Lambda^2 \right] 
\eea
and
\bea
\beta_0 &=& \frac{7\Lambda}{4 \pi r} \nonumber\\
\beta_1 &=& \frac{s^2\,\Lambda^3}{2 \pi r} + s^4 \left[ 
- \frac{5 \Lambda^3}{2 \pi r} + \pi r \right]\nonumber\\
\beta_2 &=& -\frac{5 s^4 \Lambda^3}{2 \pi r} + s^2 \left[ 
\frac{\Lambda^3 - 7\Lambda}{ \pi r} + \frac{\pi r \Lambda}{2} \right]
\nonumber\\
\beta_3 &=& \frac{4 s^6 \Lambda^3}{\pi r} + s^3 \Lambda^2 \left[ 
\frac{1 + 8\Lambda^2}{ \pi r} + \frac{\pi r}{2}(1 + 2\Lambda^2)\right]
\nonumber\\
\beta_4 &=& -\frac{2 s^8 \Lambda^3}{\pi r} - s^5 \frac{16 \Lambda^6}{\pi r}.
\eea
This expression appears to be more complicated than the one given in 
\cite{kunz}, but it can probably be transformed into it by using various 
recursion formula for the functions $F_n$, like
\be
n \ge 1 \quad \,\, x F_{n+1} = F_{n-1} - nF_n \quad {\rm{and}} \quad
\frac{dF_n}{dx} 
= F_{n+1}.
\ee

%%%---------------------------------------------------------
%%%%%%%%%%%%%%%%%%%%%%%%%%%%%%%%%%%%%%%%%%%%%%%%%%%%%%%%%%%%%%%%%%%%%%%%%%
\masection{Orthogonal case: The two-point generating function}
%%%%%%%%%%%%%%%%%%%%%%%%%%%%%%%%%%%%%%%%%%%%%%%%%%%%%%%%%%%%%%%%%%%%%%%%%%
We start from the equation
\be 
\rho^{(2)}(r_1, r_2) =  \frac{1}{2 \pi^2} \,{\hbox{Re}} \left\{ 
 \frac{\partial^2 [J_2^{+-} +J_2^{++}] } 
{\partial \eps_-(1) \,\partial \eps_-(2)}  
\biggr|_{\eps_- = \eps_+=0}\right\}, 
\label{twopt} 
\ee 
in order to compute the correlation function $\rho^{(2)}(e_1, e_2)$. Using
the fact (established in Section \ref{average}) that
\bea
J^{+-}\biggr|_{\eps_- = \eps_+} &=& 1\\
\frac{\partial J^{+-}}{\partial \eps_- (1)} &=& \pi \nu + ip \\
\frac{\partial J^{+-}}{\partial \eps_- (2)} &=& \pi \nu - ip \\
\frac{\partial^2 J^{++}}{\partial \eps_- (1) \eps_- (2)}
\biggr|_{\eps_-=\eps_+=0} &=& (\pi \nu + ip)^2, \\
\eea
we can express $\rho^{(2)}(r_1, r_2)$ as
\bea
\rho^{(2)}(r_1, r_2) &=& - \frac{1}{\lambda^2 \pi^2} p(r_1 + r_2) - 
\frac{2}{\pi^2\lambda^4} r_1 r_2 + \frac{1}{2} \nu^2 - \frac{p^2}{2\pi^2}
\nonumber\\
& & \quad - \, \frac{1}{2 d_+ d_- \lambda^2 \pi^2} 
\int D Q_+ \, DQ_- \prod_{\alpha=1}^2 
\Bigl[  Q_-(1\alpha|1\alpha) + Q_-(2\alpha|2\alpha) \Bigr]\nonumber\\
& & \, \times \, \exp\Bigl( - \frac{1}{2}\tr\bigl[ 
(Q_- - \frac{\rho_-}{\lambda} - i\beta p \lambda)^2 +
(Q_+ + \frac{i\rho_+ S}{\lambda} -\frac{ \beta p \lambda}{2})^2 
\bigr]\Bigr) \times K(q_+, q_-),\nonumber\\
\eea
where 
\be
K(q_+, q_-) = \int D\ochi \chi \exp\left(-(\ochi, \chi)\right)\,
\exp\left(\lambda \nu (e) [\frac{1}{2} A_0 + A_1 + A_2]\right);
\ee
with
\be
A_0 = i \pi \tr \left[ (q_+ + iq_-) \sigma(q_+)\right],
\ee
\be
A_1 =\int_{-\infty}^{+\infty} dt \, [\Delta(t)]^{\frac{1}{2}}\,
\left[ 1 - \frac{1}{2} \sum_{p=1}^2 \sum_{\gamma=1}^2 
\frac{q_+ (p\gamma) + i q_-(\gamma)}{t + q_+(p \gamma)}\right],
\ee
\be
A_2 = \int_{-\infty}^{+\infty} dt \, [\Delta(t)]^{\frac{1}{2}}\,
\left[ \det( \idty_2 - R)^{\frac{1}{2}} - 1\right],
\ee
and 
\be
\Delta (t) = \frac{(t - iq_-(1))^2\,(t - iq_-(2))^2}{\prod_{p=1}^2\prod_{
\gamma=1}^2 (t + q_+(p\gamma))}.
\ee
The $2 \times 2$ matrix $R$ is given by
\be
R_{\alpha \alpha'} = \frac{1}{t - iq_-(\alpha)} 
 \sum_{p=1}^2 \sum_{\gamma=1}^2 \frac{ \ochi_\alpha (p\gamma)
\chi_{\alpha'} (p\gamma)}{ t + q_+(p\gamma)}.
\ee
We recall that $q_+(p\gamma)$ and $q_-(\gamma)$ are the eigenvalues
of $Q_+$ and $Q_-$. Two eigenvalues of $Q_+$ have a positive imaginary part.
The other two have a negative imaginary part. 

$A_1$ can be expressed in terms of elliptic integrals. The same is 
true for $A_2$, once we note that 
\be
\det(\idty_2 -R) = 1 -X\, \quad {\hbox{with}} \quad  X = \tr R - \det R,
\ee
so that 
\be
\det(\idty_2 -R)^{\frac{1}{2}} - 1 = \sum_{n=1}^8 X^n c_n,
\ee
since the $X$ are Grassmannian variables [$c_n$ are constants].
One can then integrate over the Grassmannian variables, to compute
$K$. We will not reproduce this computation here, since we have not
analysed the resulting expression for the correlation function. It
can finally be noted that for higher order correlation functions,
hyperelliptic integrals appear.
%\appendix
\section*{Appendix A: Mathematical Preliminaries}
\label{appendixA}
\renewcommand{\theequation}{\mbox{A.\arabic{equation}}}
\setcounter{equation}{0}

In this appendix we list certain mathematical relations that we make use
of later in the paper.

The following Gaussian integrals play a central role in our technique for 
studying mixed systems:
\smallskip

\noindent
For a $2n \times 2n$, antisymmetric matrix $M_-$, we have the relation
\be
\int D\phi_- e^{-(\phi_-, M_- \phi_-)} = \sqrt{\det\, M_-},
\label{id1}
\ee
where $\phi_-$ denotes a vector whose elements $\phi_{-j}$ are Grassmannian 
variables, and   
$$D\phi_- := \prod_j \frac{d \phi_{-j}}{\sqrt{2}}.$$We use the convention 
that
	$$ \int d\phi_{-j} \, \phi_{-j} = {1}.$$
\smallskip

\noindent
For a symmetric matrix $M_+$, with a positive definite real part,
we have the relation
\be
\int D\phi_+ e^{-(\phi_+, M_+ \phi_+)} = \left[{\det\, M_+}\right]^{-1/2},
\label{id2}
\ee
where $\phi_+$ denotes a vector whose elements $\phi_{+j}$ are real
variables and $D\phi_+ := \prod_j (1/\sqrt{ \pi})\, d \phi_{+j}$. 
\smallskip

\noindent
A similar identity for integrals over pairs of complex variables
$\phi_{+j}, \ophi_{+j}$, which is valid for any matrix $M$ such
that $(M + M^\dagger)$ is positive definite, is
\be
\int \D\phi_+ e^{-(\ophi_+, M \phi_+)} = \left[{\det\, M}\right]^{-1},
\label{id22}
\ee
where $$ \D\phi_+ = \prod_j \frac{d\ophi_{+j}\, d\phi_{+j}}{2\pi i}.$$

The determinant of any matrix $M$ can be expressed as an integral over
Grassmannian variables as follows
\be
\int \D\phi_- e^{-(\ophi_-, M \phi_-)} = {\det\, M_-},
\label{id3}
\ee
where $\ophi_-$ and $\phi_-$ denote vectors whose elements, denoted by 
$\ophi_{-j}$ and $\phi_{-j}$, 
 are Grassmannian 
variables; the measure is given by $\D\phi_- = D\ophi_- D\phi_-$, with
$D\ophi_- := \prod_j d \psi_{-j}$ and
$D\phi_- := \prod_j d \phi_{-j}$.
\smallskip

The following identities involving determinants and traces of matrices are 
also used frequently:
Let $A$ be an $(n \times n)$ matrix  and  $B$ be an $(m \times m)$ matrix. Then
\begin{itemize}
\item{\be
\det(A \otimes B) = (\det A )^m (\det B)^n
\label{det1}
\ee}
\item{
\be
\det (A \otimes \idty_m + \idty_n \otimes B) = \prod_{\alpha= 1}^n 
\prod_{j=1}^m \left(a_\alpha + b_j\right) = 
 \prod_{\alpha= 1}^n \det \left(a_\alpha \idty_m + B\right)
= \prod_{j= 1}^m \det \left(A + b_j \idty_n\right),
\label{det2}
\ee
where $a_\alpha$ and $b_j$ denote the eigenvalues of the matrices
$A$ and $B$ respectively. The symbol $\idty_j$ is used to denote 
the $j \times j$ identity 
matrix. }
\item{If $\tr A^j = \tr B^j$ for any arbitrary integer $j$, then
\be
\det\,(\idty_n + A) = \det\,(\idty_m + B).
\label{det3}
\ee}
\end{itemize}

\section*{Appendix B: Properties of the matrix $Q_-$}
\label{appendixB}
\renewcommand{\theequation}{\mbox{B.\arabic{equation}}}
\setcounter{equation}{0}

The $2n \times 2n$ matrix $Q_-$ is self-adjoint and satisfies the property
\be 
Q_-^t = - \Upsilon_- Q_- \Upsilon_-
\label{qq-} 
\ee 
where 
$$ 
\Upsilon =  
\left( \begin{array}{lr} 0 & \idty_n \\
		  - \idty_n & 0 
\end{array} \right)
$$ 
We will show that such matrices have doubly degenerate eigenvalues
and can be diagonalised by unitary matrices, which also 
belong to the symplectic group. More precisely
\be
Q_- = U q_- U^\dagger,
\label{a3}
\ee
where
\be
U U^\dagger = U^\dagger U = 1 
\ee
\be
U^t \Upsilon_- U = \Upsilon_-
\label{five}
\ee
and $q_-$ is the diagonal matrix of eigenvalues of $Q_-$.
Consider the spectral decomposition of $Q_-$:
\be
Q_- = \sum_{j=1}^n \lambda_j P_j.
\ee
Then the projectors $P_j$ also have the property \reff{qq-}, which implies
that 
$$P_j (1\alpha|1 \beta) = P_j (2\beta|2 \alpha).$$
hence,
\be
\tr P_j = 2\, \sum_{\alpha = 1}^n P_j (1\alpha|1 \alpha), 
\ee
which implies that for almost all $Q_-$ (with respect to the Lebesgue
measure), $\Tr P_j =2$, and, therefore, the eigenvalues of $Q_-$ are
doubly degenerate.

On the other hand, \reff{qq-} and \reff{a3} can be expressed as
\be
M q_- =  q_- M,
\label{sev}
\ee
where 
\be
M = U^\dagger \Upsilon_- (U^\dagger)^t.
\label{eit}\ee 
However,
\be
M^t = -M \quad \quad {\rm{and}} \quad \quad M^\dagger M = \idty_{2n},
\ee
and from \reff{sev} it follows that 
\be
M(p \alpha | q \beta) = \delta_{\alpha, \beta} \, e^{i \psi_\alpha}\, [
\delta_{p,1} \delta_{q,2} - \delta_{p,2} \delta_{q,1}]
\label{nine}
\ee
Noting, however, that $U$ can be replaced by $U e^{i \phi}$ in \reff{a3},
where $e^{i \phi}$ is a diagonal matrix, we see from \reff{eit} that we can choose 
$ \psi_\alpha = 0$ in \reff{nine}. In other words 
$$M = \Upsilon_-$$
which gives the desired property \reff{five} for $U$.  

\vspace{1.5truecm}

{%\renewcommand{\baselinestretch}{1.2}\large\normalsize
\def\thebibliography#1{\noindent
{\large\bf References} \par \list
{\arabic{enumi}.}{\settowidth\labelwidth{[#1]} \leftmargin\labelwidth
\advance\leftmargin\labelsep \itemsep=0pt \usecounter{enumi}}
\def\newblock{\hskip .11em plus .33em minus -.07em} \sloppy
\sfcode`\.=1000\relax}
}
\end{document}